\documentclass[10pt]{article}
\usepackage[OE]{express}

\def\ao{Appl. Opt. }

\def\apl{Appl. Phys. Lett. }

\def\josa{ J. Opt. Soc. Am. }
\def\josaa{ J. Opt. Soc. Am. A }

\def\opex{Opt. Express }

\def\aj{Astron. J. }

\def\aap{Astron. Astrophys. }
\def\aaps{Astron. Astrophys. }
\def\mnras{Mon. Not. R. Astron. Soc. }
\def\pasp{Publ. Astron. Soc. Pac. }
\def\araa{Annu. Rev. Astron. Astrophys. }
\def\apjs{Astrophys. J. Supp. }

\def\procspie{Proc. SPIE }

\newcommand{\be}{\begin{center}}
\newcommand{\ee}{\end{center}}
\newcommand{\beN}{\begin{center}}
\newcommand{\eeN}{\end{center}}
\newcommand{\bea}{\begin{eqnarray}}
\newcommand{\eea}{\end{eqnarray}}
\newcommand{\beaN}{\begin{eqnarray*}}
\newcommand{\eeaN}{\end{eqnarray*}}

\newcommand{\Dlambda}{{\Delta\lambda}}
\newcommand{\dlambda}{{\delta\lambda}}
\newcommand{\Nlambda}{{N_\lambda}}

\newcommand{\SR}{\mathcal{R}} %Spectral Resolution
\newcommand{\St}{\mathcal{S}} %Strehl ratio

\begin{document}
\title{Spectrographs for astrophotonics}

%% OE
\author{N. Blind, \authormark{1,2,*}  E. Le Coarer,\authormark{2} P. Kern,\authormark{2} S. Gousset\authormark{2}}
\address{\authormark{1}Geneva Observatory, University of Geneva, 51, ch. des Maillettes, CH-1290 Versoix, Switzerland\\
\authormark{2}UJF-Grenoble 1/CNRS-INSU, Institut de Plan\'etologie et d'Astrophysique de Grenoble (IPAG) UMR 5274, Grenoble, France}

\email{\authormark{*}nicolas.blind@unige.ch} %% email address is required

%%%%%%%%%%%%%%%%%%%%%% Front pages %%%%%%%%%%%%%%%%%%%%%%

%\abstract{Context}{Aims}{Methods}{Results}{Conclusions}
\begin{abstract}
The next generation of Extremely Large Telescopes (ELT), with diameters up to 39 meters, is planned to begin operation in the next decade and promises new challenges in the development of instruments since the instrument size increases in proportion to the telescope diameter $D$, and the cost as $D^{2}$ or faster. The growing field of astrophotonics (the use of photonic technologies in astronomy) could solve this problem by allowing mass production of fully integrated and robust instruments combining various optical functions, with the potential to reduce the size, complexity and cost of instruments. Astrophotonics allows for a broad range of new optical functions, with applications ranging from sky background filtering, high spatial and spectral resolution imaging and spectroscopy. In this paper, we want to provide astronomers with valuable keys to understand how photonics solutions can be implemented (or not) according to the foreseen applications. The paper introduces first key concepts linked to the characteristics of photonics technologies, placed in the framework of astronomy and spectroscopy. We then describe a series of merit criteria that help us determine the potential of a given micro-spectrograph technology for astronomy applications, and then take an inventory of the recent developments in integrated micro-spectrographs with potential for astronomy. We finally compare their performance, to finally draw a map of typical science requirements and pin the identified integrated technologies on it. We finally emphasize the necessary developments that must support micro-spectrograph in the coming years.
\end{abstract}


%\bibliographystyle{osajnl}
%%\bibpunct{[}{]}{,}{n}{}{;}
%\bibliography{/Users/nblind/Travail/Papers/biblio}   % BiTex files

\section{Introduction} % (fold)
\label{part:introduction}

Astronomy deals with the study of celestial sources through the analysis of their light. Since the start of modern astronomy, spectroscopy has played a major role with applications ranging from the study of celestial dynamics, chemical properties,  Universe dynamics, and more recently the search and characterisation for extrasolar planets. The next generation of Extremely Large Telescopes (E-ELT \cite{mcpherson_2012a}; TMT \cite{nelson_2008a}; GMT \cite{johns_2012a}) is planned to begin operation in the next decade and promises significant breakthroughs in our understanding of the Universe, as well as new challenges on the instrumentation side. In particular, the linear size of their instruments will increase in proportion of the telescope diameter $D$, while their cost increases at least as $D^2$ \cite{bland_2006a} or faster. This leads to instruments whose cost are comparable to 8m class telescopes \cite{russel_2004a, allington_2010b}.

Astrophotonics is a quickly developing research and development field that lies at the interface of astronomy and photonics \cite{bland_2012a}. The first application of astrophotonics that demonstrated a significant breakthrough was long baseline optical interferometry (LBOI) with the use of single-mode fibers \cite{coudeduforesto_1997b} and integrated beam combiners \cite{malbet_1999a, berger_2001a, lebouquin_2004a}. Almost 20 years after this pioneering work, 4-telescope integrated beam-combiners and various photonics functions (fiber delay lines and polarization control in single-mode fibers, and photonic metrology system) allow an instrument like GRAVITY \cite{eisenhauer_2017a} to provide milli-arcsecond spectro-imaging in K-band and micro-arcsecond astrometry capabilities at unprecedented levels of sensitivity in LBOI.

Astrophotonicss developments are now carried on in various fields, with applications in sky OH-lines suppression \cite{bland_2009a}, high angular resolution spectro-imaging via pupil remapping \cite{jovanovic_2012a, huby_2012a}, beam reformatting for integral field-spectroscopy  \cite{harris_2014a, maclachlan_2015a}, mode conversion with photonic lanterns \cite{horton_2014a}, and generation of frequency comb or Fabry-P\'erot cavities \cite{ellis_2012a}. Although these devices are greatly inspired by telecommunication developments, astrophotonics devices also have to answer different constraints, in particular very high throughput over much broader spectral range, of a decade or more. Astrophotonics is therefore an active field of research and development, with already a few successful instruments using several sub-components, like PIONIER \cite{lebouquin_2011a} and GRAVITY \cite{eisenhauer_2017a} for optical interferometry, or GNOSIS \cite{trinh_2013a} for integral field spectroscopy.

In this paper, we focus on the developments of integrated spectrographs with a view to achieving the following goals:
\begin{itemize}
\item to conceive more compact and simpler instruments than classical bulk optics systems;
\item to minimize useless pixels thanks to a better/easier arrangement of light onto the detector (conversely to, e.g., lenslet and masked multi-object spectrographs like TIGER \cite{bacon_1988a} or LUCI \cite{buschkamp_2010a});
\item to improve the stability of instruments by drastically limiting the number of individual and/or moving optical parts, as well as reducing the overall dimensions of the instrument.
\end{itemize}
Astrophotonic spectrographs can potentially provide the solution to the continuous increase in complexity of future optical/infra-red ground-based spectrographs, while reducing the instruments size, as well as their cost through mass production capabilities. The compactness of these instruments is one of the most obvious advantages in the context of space missions, and these have already found applications on rockets, balloons, drones \cite{fogarty_2012a, betters_2012a}, or even micro-satellites \cite{gousset_2016a, lecoarer_2016a}.

In this paper, we want to provide astronomers with valuable keys to understand how photonics solutions can be implemented (or not) according to the foreseen applications. The paper is organized as follow. In Section~\ref{part:group_spectro}, we recall some basic principles of spectroscopy, placed in the framework of astrophotonics. Section~\ref{part:optical_etendue} focuses on the notion of optical \'etendue and its implications on photonics spectrograph. In Section~\ref{part:identified_technos}, we define a series of merit criteria for astrophotonics allowing then to select and compare a non-exhaustive list of technologies identified with potential for astronomy. Performances of the different families of integrated spectrographs are finally compared in Section~\ref{part:signal_to_noise}, in the case of typical astrophysical sources and observing conditions. In Section~\ref{part:conclusion}, we summarize our reasons to believe that integrated spectrographs will be more widely used in the near future, but must also come along with the development of other subsystems, in particular adaptive optics and new detectors technologies.

%%%%%%%%%%%%%%%%%%%%%%%%%%%%%%%%%%%%%%%%%%%%%%%%%%%%%%%%%%%%
%%%%%%%%%%%%%%%%%%%%%%%%%%%%%%%%%%%%%%%%%%%%%%%%%%%%%%%%%%%%
%%%%%%%%%%%%%%%%%%%%%%%%%%%%%%%%%%%%%%%%%%%%%%%%%%%%%%%%%%%%

\section{Groups of spectrometers} % (fold)
\label{part:group_spectro}

\begin{figure}[t!]
\centering
\includegraphics[width=.98\textwidth]{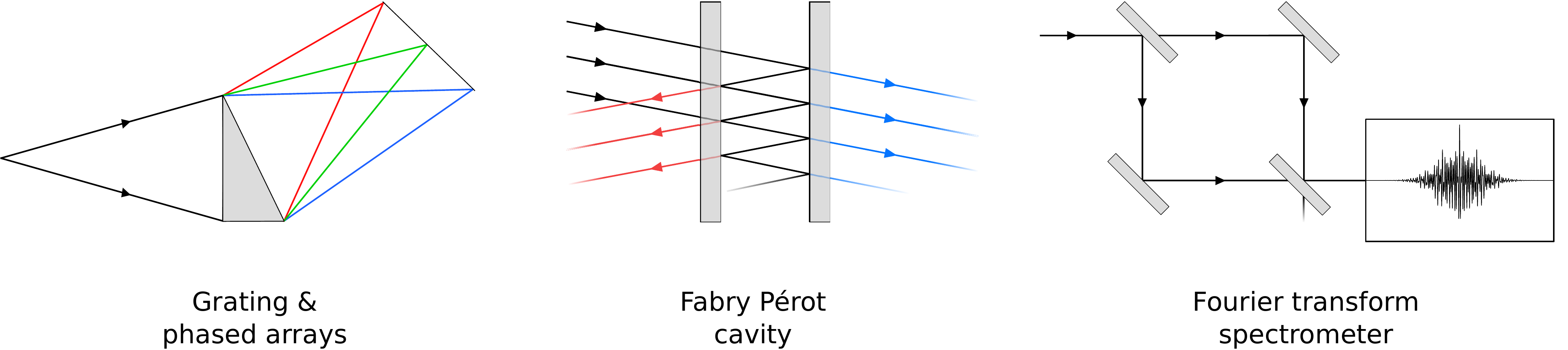}
\caption{Schematics of the three groups of spectrometers.}
\label{fig:groups_spectrometer}
\end{figure}

The Wiener-Khinchin theorem states that the power spectral density of a stationary random process is the Fourier transform of the corresponding autocorrelation function. It is the basis of spectrometry, and can be summarized as an autocorrelation of light propagation in time. The maximum spectral resolution is theoretically determined by the maximum optical delay $\Delta$ applied between parts of the wavefront, such as: $\SR$ = 2 $\Delta/\lambda$. We distinguish three groups of spectrometer in this paper [Fig.~\ref{fig:groups_spectrometer}].

%%%%%%%%%%%%%%%%%%%%%%%%%%%%%%%%%
%%%%%%%%%%%%%%%%%%%%%%%%%%%%%%%%%

\subsection{Grating and phased arrays} % (fold)
\label{part:dispersion}
Gratings are currently the most common type of spectrograph and also the most sensitive. Gratings, prisms, or any phased array of facets or waveguides permit to quasi continuously delay part of the wavefront. Each wavelength then interferes constructively on a different position on the detector, allowing to observe the spectral power density (i.e.~the spectrum) of the signal. The dispersion and interference process operated by the device actually corresponds to a (noise-free) temporal Fourier Transform (FT) of the incoming wavefront. Although grating spectrometers are the most efficient in terms of Signal-to-Noise Ratio (SNR), they suffer of a restricted optical \'etendue.

% subsubsection dispersion (end)

\subsection{Fabry-P\'erots and cavities} % (fold)
\label{part:filters}
In Fabry-P\'erot or interferential filters, the wavefront is progressively delayed over reflective surfaces before being recombined. A few layer arrangement allows the elimination of unwanted spectral bands, or to multiply the spectra by a comb which is separated by remaining optics. 
Efficient filters are based on interferential processes which achieve an in-line interferential reconstruction of delayed parts of the wavefront (reflective coating, Fabry-P\'erot, Lippmann-Bragg variable index distribution, polarizing materials in Lyot filters, etc.). In bulk devices, most of the spectrum is reflected and generally lost. This makes such filters rather inefficient if the reflected light carries useful information. On the other hand, a Fabry-P\'erot instrument is very efficient for spatially and spectrally observing very extended sources dominated by their emission lines (e.g. for studying kinematics distribution in galaxies \cite{bland_1989a}).
A few photonics solutions offer the possibility to cascade such filters, enabling the design of broadband, high resolution spectrometers with efficiency equivalent to gratings.

% subsubsection filters (end)

\subsection{Fourier transform spectrometers} % (fold)
\label{part:fourier}

In Fourier Transform Spectrometers (FTS), the wavefront is separated in two parts, before being recombined with a variable delay forming interference fringes. Performing a digital Fourier Transform of these latters, the  spectrum is reconstructed. In general, FTS instruments are not very sensitive, as the detection noise applies to each fringe sample, and consequently pollutes every spectral element after the FT operation. In domain where detectors are noisy, rare or expensive, FTS benefits however of the multiplex advantage \cite{fellgett_1958a}. For instance, an imaging FTS can be very simple, efficient and compact to observe HII regions \cite{maillard_2013a}.

% subsubsection fourier (end)

\subsection{Direct detection} % (fold)
\label{part:direct_detection}

Different detector technologies are able to directly measure the energy of incident photons. Table~\ref{tab:detector} presents a list of few technologies in a research and development step. These detectors can only provide a low resolving power up to $\sim$100. They can be used in many domain as low spectral resolution spectro-imagers, but could also replace the cross-disperser for high resolution spectrographs, hence keeping the instrument compactness and simplicity, as well as requiring 20 to 100 times less pixels. The MKID (Magnetic Kinetic Inductance Detector) technology is already in operation in ARCONS-2024 \cite{mazin_2013a}, and few other instruments are also studied (KRAKENS \cite{mazin_2015b}, MEC for exoplanet imaging \cite{mazin_2015a}, or a focal plane camera for the LSST \cite{mazin_2012a}), with goal to achieve 60kpix detectors by 2020. Since they cannot offer medium and high spectral resolution, they are not discussed in the following. The interested reader can find a more detailed review of these technologies in \cite{eisenhauer_2015a}.

% subsubsection direct detection (end)

 %%%%%%%%%%%%%%%%%%%%%%%%%%%%%%%%%
\begin{table*}[t]
\centering
\caption{List of different energy-sensitive detectors under development. \label{tab:detector}}
\begin{tabular}{llll}
\hline \hline
Technology & $\Delta E$ & $\SR$ = $\Delta E/ E$ & Ref. \\
\hline
Transition Edge Superconductor & $\geq 0.05  eV$ & $20$ & \cite{burney_2006a}\\
Superconducting Tunneling Junction & $\geq 0.01 eV$ & 100 & \cite{jorel_2005a}\\
Magnetic Kinetics Inductance & $\geq 0.01  eV$ &100 & \cite{mazin_2013a}\\ 
Quantum Dots (PbS) &  $\geq 0.3  eV$ & 5 & \cite{konstantatos_2007a}\\
\hline
\end{tabular}
\end{table*}
%%%%%%%%%%%%%%%%%%%%%%%%%%%%%%%%

%%%%%%%%%%%%%%%%%%%%%%%%%%%%%%%%%%%%%%%%%%%%%%%%%%%%%%%%%%%%
%%%%%%%%%%%%%%%%%%%%%%%%%%%%%%%%%%%%%%%%%%%%%%%%%%%%%%%%%%%%
%%%%%%%%%%%%%%%%%%%%%%%%%%%%%%%%%%%%%%%%%%%%%%%%%%%%%%%%%%%%
\section{Optical \'etendue} % (fold)
\label{part:optical_etendue}

\subsection{Source \'etendue}

Because of the \'etendue conservation law, the larger is the observed solid angle and/or the telescope size, the larger must be the spectrograph, very often equipped with gratings of theoretical resolution factors few above the need of the instrument. The notion of optical \'etendue is therefore a key parameter which is at the heart of this paper and strongly influences performance of integrated spectrographs. In order to work in the most general context in the following, we will consider a {\it normalized Optical Etendue $S\omega$}, defined independently of the wavelength:
\begin{equation}
		S\omega = S\Omega/\lambda^2,
\end{equation}
\noindent where $S$ is the telescope collecting area, $\Omega$ the object solid angle, normalized by the squared wavelength $\lambda^2$. For a telescope at the diffraction limit, the normalized optical \'etendue of a spatial resolution element is close to unity: $S\omega \sim 1.2$. Normalised optical \'etendue is therefore a measure of the spatial information content of the beam (whether or not it is used), which varies with wavelength for a given field of view and telescope aperture. In the following we distinguish two types of sources depending on their extent as defined by their normalised solid angle $S\omega$:
\begin{description}
\item[\bf Point source] -- A point source is by definition a source for which $S\omega\sim1.2$. This is a spatially unresolved source observed with a telescope at the limit of diffraction (e.g.~space telescope or ground-based one with an extreme Adaptive Optics (AO)).

\item[\bf Extended source] -- On the contrary an extended source is so that $S\omega>1.2$. This is the case of a spatially resolved source (typically galaxies or nebulae) or of an unresolved source observed under a turbulent atmosphere (e.g.~a star whose light is spread over a speckle pattern if low order or no AO correction is applied).
\end{description}
%

%As long as a source is smaller than the telescope PSF (either turbulent, partially corrected by AO or turbulence free), increasing the telescope diameter increases the quantity of light it collects (i.e.~its sensitivity), without any impact on its dimension.
For ground-based telescopes, the advantage of an AO correction is twofold: it reduces the full instrumentation size, and also reduces the background noise due to the telescope and the sky thermal emissions (which rapidly dominate from the near-IR at $\lambda \ge$2.0$\mu m$; Table~\ref{tab:background}), or the telluric OH emission lines that can rapidly saturate areas of the detector.

\smallskip
When the source is larger than the telescope PSF (which can be either turbulent, AO corrected or diffraction limited), we can consider two transverse use cases:
\begin{description}
\item[Spectro-imaging] -- If the source must be spatially sampled at the Shannon frequency, there is no sensitivity gain anymore by increasing the telescope size since the \'etendue of each spatial element does not increase anymore.  The gain is on spatial information.

\item[Faint source] --  For faint and extended sources, priority can be given to the spectral information instead of spatial one. The telescope is then used as a photon collector and its sensitivity is increased in proportion of the telescope collecting area.
\end{description}
\noindent Depending on the science case and quantity of interest (spatial or spectral), a compromise is made between maximum sensitivity and full spatial sampling.

\subsection{Instrument \'etendue and number of modes}
\label{part:modal_purity}

In the framework of astrophotonics, micro-spectrographs mostly rely on the use of waveguides to propagate and produce the light spectrum. The electric field propagating in a waveguide can be decomposed on the basis of a limited number of orthogonal modes, whose number and shape varies as a function of wavelength and of the waveguide properties (optical index, geometry, etc.). This number can vary from 1 for the so-called single-mode waveguide, and up to few thousands for visible instruments with fibers size matched to the local seeing size.

\subsubsection{Single-mode instruments}

Waveguides are called Single-Mode (SM) when only the fundamental, nearly gaussian mode can propagate. SM waveguides have an optical \'etendue $S\Omega|_{SM}$ = $\lambda^2$, or $S\omega|_{SM}$ = 1.

The coupling efficiency $\rho$ of a beam into a waveguide depends on the matching between the intensity profile of the injected beam and the waveguide modes. Coupling efficiency in a SM fibers can not be computed with standard ray tracing approximation, but requires instead to compute the convolution between the incident electromagnetic field and the fiber fundamental mode, for each polarization state. For an unobstructed diffraction limited circular telescope, a maximum coupling efficiency $\rho_0 \sim 0.82$ \cite{shaklan_1988} is achieved. For obstructions of 20\%, $\rho_0$ drops to $\sim 70$\% \cite{ruilier_2001a}. By using Phase Induced Amplitude apodization masks \cite{guyon_2004a}, it is actually possible to remap the rays to apodize the telescope pupil amplitude and eliminate the telescope central obscuration in a way best matching the single-mode fiber unique mode \cite{jovanovic_2016b}. Therefore, such apodizers can overcome the limit presented in \cite{shaklan_1988} and allow up to 100\% coupling in SMF. Practical values of up to 65\% have already been reported \cite{jovanovic_2016b}.

In the case of an extended source, coupling of off-axis field drops rapidly, leading to low efficiency and lower weighting of these part of the spectrum.

%%%%%%%%%%%%%%%%%%%%%%%%%%%%%%%%%%%%%%%%%%%%%%
\begin{table*}[t]
\centering
%\begin{minipage}{0.7\textwidth}
\centering \caption[Typical sky background flux in visible and IR. ]{Typical sky background flux in visible and IR \cite{patat_2003a, cuby_2000a}. \label{tab:background}}
\begin{tabular}{llrrrrrrr}
\hline \hline
Band && V & I & J & H & Ks & L & M\\
\hline
$B_\lambda$ &[mag.arsec$^{-2}$] & 21.6 & 19.6 & 16.5 &14.4 & 13.0 & 3.9 & 1.2 \\
$B_\lambda$ for $S\omega$= 1 &[ph.s$^{-1}$] & 0.2 & 4 & $77$ & $730$ & $2\cdot 10^3$ & $9\cdot 10^6$ & $8\cdot 10^7$\\
\hline
\end{tabular}
%\end{minipage}
\end{table*}
%%%%%%%%%%%%%%%%%%%%%%%%%%%%%%%%%%%%%%%%%%%%%

\subsubsection{Multi-mode instruments}

Waveguides are multi-mode (MM) when two or more modes can propagate. In this respect, we consider bulk optics as MM. MM devices have consequently a higher optical \'etendue and can efficiently couple light from extended sources. For a number of supported modes above $\sim$ 100, geometric optics explains well the behavior of MM waveguides (private com.), like bend losses, far- and near-field scrambling properties \cite{chazelas_2010b}. $\mathrm{N_{mode}} < 100$ is the regime of the few-mode fibers, where geometrics optics does not apply anymore: coupling calculations are similar to the single-mode case, by computing the convolution of the incident electromagnetic field to every individual modes in the fiber. Doing so, \cite{horton_2007a} shows that few-mode fibers have an optical \'etendue close from the geometrical optics expectations, even for bi-mode fibers. However, the mismatch between the telescope PSF and the few modes leads to lower maximal coupling efficiency of $\sim$80\% in the bi-mode case, slowly increasing to $\sim$90\% for a 3 times larger fiber core with $\mathrm{N_{mode}}$ = 25, again in the case of an unobstructed telescope. Similarly to single-mode fibers, in this regime, telescope central obstruction reduces maximum coupling, to around 60\% for a 30\% obstruction (private com.). Explaining the internal behavior of few-mode fibers also requires more complex simulation tools like finite element models with proper understanding and modeling of fiber imperfections and its impact on mode-to-mode couplings.

Considering a circular, graded MM waveguide with a core of radius $r$ and a numerical aperture $NA$, its {\it normalized} optical \'etendue is therefore:
\begin{equation}
	S\omega|_{MM} = \left( \dfrac{\pi r NA}{4\lambda}\right)^2,
\end{equation}
while the number of modes is closely approximated by:
\begin{equation}
	M = \frac{V_{fiber}^2}{4},
\end{equation}
\noindent with $V_{fiber}$ = $\pi r NA/\lambda$. Those two relations hold even for M=2 \cite{horton_2007a}. The number of modes that a MM waveguide can propagate is then roughly equal to its normalized optical \'etendue:
\begin{equation}
	 S\omega|_{MM} = \frac{\pi}{4}  M \sim 0.8  M\; .
\end{equation}
These relations are not valid for square and rectangular waveguides. The asymmetry of the waveguides makes it naturally birefringent, with a different number of modes transported for each polarization.

Note that generally, the acceptance cone of MM waveguides are underfilled. Because of waveguide imperfections, strains and connector stresses, the output cone is larger than the one injected (i.e. the output \'etendue is larger than the input one). This effect known as Focal Ratio Degradation (FRD) ultimately results in a transmission loss if the instrument has been designed according to the source \'etendue. FRD losses can be reduced by oversizing the spectrograph, although not desirable. On the other hand, if the fiber cone is filled, FRD is zero by definition, but fiber imperfections and bends lead to higher losses into evanescent modes of the fiber.

\subsubsection{Multiplexed single-mode instruments}
Multiplexed single mode (MSM) spectrograph is a class of spectrographs specific to astrophotonics and enabled by the mass production capabilities of photonic devices. They can be fed with MM-to-SM mode converters like the photonic lanterns \cite{leon-saval_2005a, noordegraaf_2009a}. They consist of $N_g$ SM instruments -- each accepting a unit normalized optical \'etendue $S\omega|_{SM}$ = 1 --, which combine the advantage of SM devices (very high spectral resolution and stability) and of MM devices (large optical \'etendue, proportional to the number of input SM waveguides), so that:
\begin{equation}
S\omega|_{MSM} = N_g\;.
\end{equation}
\noindent $N_g$ SM spectrographs can sample the object under study, and measure the same spectrum. The read-out noise being also multiplied in proportion, such an instrument has limited sensitivity for faint targets, especially with ELTs. FTS like SWIFTS \cite{lecoarer_2007a} can however take advantage of the large optical \'etendue of the source to increase the spectral resolution and/or wavelength coverage by sampling a different part of the interferogram on different waveguides

\subsection{Limits in spectral resolution} \label{part:jacquinot}

\subsubsection{Source solid angle}

The Van Cittert and Wiener theorems are mutually constrained by the Jacquinot criterion which states that the Field Of View of any spectrometer must be limited if a high spectral resolution $\SR$ is desired. For FP and FTS, the limit is $\Omega < 2\pi/\SR$, while for gratings in Littrow condition and incidence angle $\beta$, $\Omega < \beta / \SR$ \cite{jacquinot_1954a}. Hence, for bulk instruments, a grating has a field of view 10 to 100 times smaller than FP and FTS.

\subsubsection{Quality factor}

Spectroscopy relies on the ability to perform a nearly perfect temporal autocorrelation and Fourier Transform of the incoming light. Hence, for gratings and waveguide phased arrays, any imprecision in applying the correct amount of delay (through glass inhomogeneities, facet surface errors, etc.) will result in a distorted spectrum with reduced resolution. In Fabry-P\'erot cavities, the delay is applied between the two plates, so that the limiting factor is their surface quality and misalignment errors. In FTS, errors on the different delays generate numerous artifacts in the reconstructed spectrum, resulting in a limited resolution and reconstruction fidelity. The advantage of {\it static} FTS (with no moving parts) over all other spectrographs is that the delays can be very precisely calibrated and taken into account in a finely tuned, non linear FT. This is the only system where the spectral resolution can be preserved in presence of imperfections.

\subsubsection{Modal dispersion and modal noise} \label{part:modicity_limitation}

In the framework of astrophotonics, the source solid angle plays a smaller role since the light is collected through a big fiber or a fiber bundle, which is then split towards a series of spectrographs. Spectral resolution can then be limited at the spectrograph level by modal dispersion, if the waveguides are multi-mode. The optical path traveled by each mode can differ by a few percents as imposed by the optical index contrast between the waveguide core and its cladding, meaning that each mode will produce its own spectrum at a different position on the detector. Doing so, spectral resolution can be limited to a few 1000.

MM devices are also subject to modal noise. Modal noise is due to external factors like instability of the PSF injection (tip-tilt error), temperature changes and variations in fiber routing (e.g.~caused by telescope tracking). Modal noise arises from variations of the modal configuration and modal dispersion in MM waveguides that lead to a changing instrumental response. Scrambling strategies are applied in order to lose this information and get an instrumental response as stable as possible under any possible conditions. Modal is inverse proportional to the number of modes, i.e. it increases with smaller fiber (until SM case is reached) and longer wavelengths.

\subsubsection{Birefringence}

Equivalently, if the waveguide is birefringent, the spectral resolution of the instrument can be dramatically limited. Delays will be different for both polarizations, leading to superimposed spectra of the same signal at different positions onto the detector for a grating. In the case of an FTS, two interference patterns of different frequencies are mixed, resulting in a beating of the fringe contrast. $\delta n$ being the difference of optical index between the ordinary and extraordinary axes, the resolution is typically limited to $\SR < n/\delta n$. In silica,  $\delta n$ = $10^{-4}$ leads to $\SR < 15000$, so that high resolution integrated spectrographs probably require to split polarizations.

\subsection{The glass immersion and photonics advantages}

Since the spectral resolution is determined by the maximum {\it optical} delay $\Delta$ applied within the spectrometer, delaying in a glass of high optical index like silicon ($n \sim 3.4$ in NIR) or Germanium ($n \sim 2.4$ in NIR) rather than in air reduces the size of the dispersing element by an equivalent factor. Developments on immersion gratings are ongoing for the next generation of spectrographs for the ELTs \cite{rodenhuis_2015a}, with already some demonstrators on sky, like IGRINS \cite{park_2014a, mace_2016a}. This concept is of course equally applicable -and applied- to integrated spectrographs. Working in a medium of high optical index comes with two additional advantages:
\begin{itemize}
\item Increased field of view at given resolution - With an immersed device, the source is also seen with a smaller solid angle $\Omega/n^2$, meaning that the field of view is increased by a factor $n^2$ at constant spectral resolution. This is the case of FTS spectrometers like $\mu$SPOC \cite{gillard_2012a} or SWIFTS-LA \cite{lecoarer_2014a}.
\item Increased compactness - For purely integrated spectropgraphs, having a high index contrast $\Delta n \sim$ 20-40\% like with Silicon-Nitride-On-Insulator platform confines much more strongly the modes into the waveguide than standard Silicon-On-Silica technologies ($\Delta n \sim 0.7$\%). This appears of particular use for devices like Array Waveguide Grating (AWG; Section~\ref{part:awg_and_planar_gratings}) to reduce the bend radius of the phased array section as well as limit cross-couplings between each waveguide \cite{cheben_2007a, fernando_2014a}. In theory, a near infrared spectrograph of resolution 100000 drawn on silicon could feet on a substrate of 3cm length and few mm thickness.
\end{itemize}
On the other hand, high index also present drawbacks:
\begin{itemize}
\item High index contrast involves smaller modes and higher numerical aperture, making efficient coupling more difficult;
\item Fresnel losses at interfaces are higher, although anti-reflection coatings of the polished fiber can be applied and significantly reduce the losses.
\end{itemize}

%%%%%%%%%%%%%%%%%%%%%%%%%%%%%%%%%%%%%%%%%%%%%%%%%%%%%%%%%%%%
%%%%%%%%%%%%%%%%%%%%%%%%%%%%%%%%%%%%%%%%%%%%%%%%%%%%%%%%%%%%
%%%%%%%%%%%%%%%%%%%%%%%%%%%%%%%%%%%%%%%%%%%%%%%%%%%%%%%%%%%%
\section{Identified integrated technologies}
\label{part:identified_technos}

This section presents a non-exhaustive overview of different integrated spectrograph technologies in development in the field of telecommunications, bio- and environmental sensing, or specifically for astronomy. These solutions have been selected and identified as relevant in the framework of astrophotonics on the basis of a series of merit criteria.

\subsection{\bf Merit criteria}
\label{part:merit_criteria of integrated spectrometers}

\paragraph{Multiplexing capability} -- With astrophotonics, there are two different ways to access to multiplexing capabilities:\smallskip

\noindent {$\bullet$ In-chip multiplexing} -- Certain technologies allow to measure several different spectra within a single integrated spectrograph (up to 30; Section~\ref{part:side_holographic_dispersion})). The cost in terms of volume can be high since it generally requires a cross-dispersion (bulk) stage to disentangle spectra. Energy-sensitive detectors could be a solution to preserve compactness. \smallskip

\noindent {$\bullet$ 3D multiplexing} -- Depending on the technology, it is more or less easy to increase and integrate together a large number of spectrographs. For instance, multiplexing capability of {\it semi}-integrated spectrographs (containing bulk optics) is limited, since the compact size makes the precise alignment of small optical components very difficult. On the other hand, fully integrated planar technologies allow to easily stack tenths of spectrographs in a very compact and stable way.
\paragraph{Operability} -- Instruments can necessitate to be finely tuned or calibrated because of their sensitivity to any position change along its use (e.g.~Fabry-P\'erot are very sensitive to altitude evolution). Fully integrated spectrographs have the advantage to have no moving parts and to provide more stable set-up as it has been demonstrated numerous times in optical interferometry for instance \cite{berger_2003, lebouquin_2004a}. The assembly of the different subsystems is precisely done at manufacturing time and the system size remains small, allowing a secure packaging. We can also include critical assembly steps such as coupling light to the fibers or to micro-lenses.

`A temperature tuning may be necessary for some technologies, depending on the requirements on spectral resolution, stability, etc. As an example, the ring resonators (Section~\ref{part:filter_ring_resonator}) tend to show a rather low sensitivity (50mK temperature stability corresponding to 1m/s), although it is not excluded it is a technology dependent factor. On the other hand, the compactness of integrated devices may also to reach such degrees of stability: the RHEA experience \cite{feger_2016a} shows that a shoe-box bulk spectrograph can be easily stabilized to better than 10mK to reach 1m/s stability.

\paragraph{Spectral Filling Factor $F_F$} -- A spectrum can be considered as a limited bandwidth signal made of $k$ spread lines of equivalent width $W_i$ in a wide spectral range $\Dlambda$. The spectral filling factor $F_F$ characterizes then the quantity of information contained in the spectral bandwidth:
\begin{equation}
F_F = \frac{\sum_i^k W_i }{\Dlambda}\;.
\end{equation}
It is a property inherent to the source, but different kind of spectrographs will more or less well match this constraint. For instance, for small filling factor (e.g.~nebulae), Fabry-Perot or FTS are in general favored by allowing the selection of a few specific lines. For important filling factors gratings are in general preferred, like in the case of radial velocities measurements, where $F_F$ can be as high as 60\% \cite{lovis_2006a}.

\begin{figure}[b!]
\centering
\includegraphics[width=0.68\textwidth]{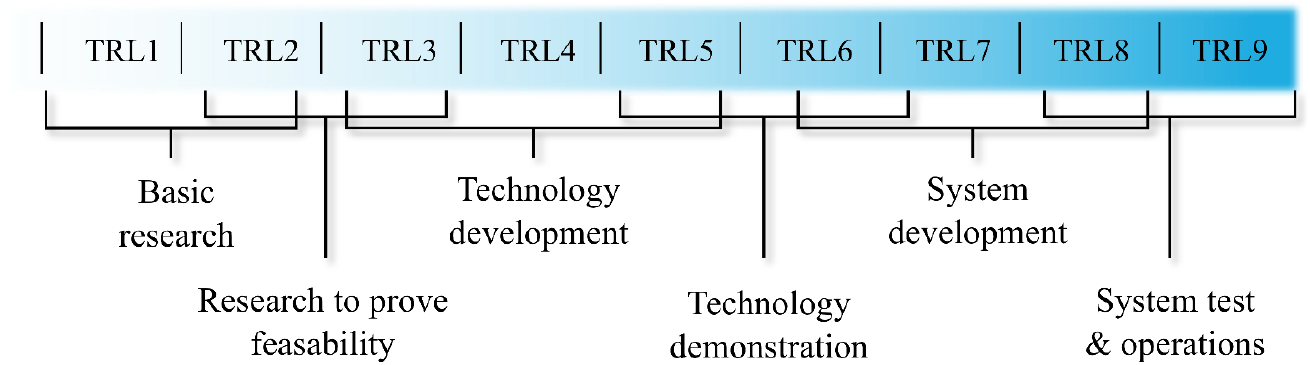}
\caption{Schematic presentaion of NASA TRL levels. \label{fig:TRL}}
\end{figure}

\paragraph{Technology Readiness Level} -- The different integrated spectrographs considered in this document are at different stages of development. Although it is not totally adapted to ground-based instruments, we use the NASA Technology Readiness Level (TRL) \cite{mankins_1995a}. This scale has also been adopted by ESO for the development track of the E-ELT:

\begin{itemize}
\item{\bf TRL 1}  -- Basic principles observed and reported.
\item{\bf TRL 2} -- Technology concept and/or application formulated.
\item{\bf TRL 3} -- Analytic and experimental critical function and/or characteristic proof-of-concept.
\item{\bf TRL 4} -- Component and/or breadboard validation in laboratory environment.
\item{\bf TRL 5} -- Component and/or breadboard validation in relevant environment.
\item{\bf TRL 6} -- System/subsystem model or prototype demonstration in a relevant environment.
\item{\bf TRL 7} -- System prototype demonstration in a space environment.
\item{\bf TRL 8} -- Actual system completed through test and demonstration (ground or space).
\item{\bf TRL 9} -- Actual system proven through successful (mission) operations.
\end{itemize}

In this paper, we use this scale to assess the maturity of the different spectrographs we identified, and it can be read in a more simplified way as presented in Fig.~\ref{fig:TRL}. In particular in this work, identified technologies with TRL $\leq 4$ are in early development phases, with, in addition different applications and requirements in mind by the authors of the original papers. For such technologies, it is difficult to assess the limits and challenges that we may face on later demonstration and integration stages.

%\subsection33{Technologies}
%\label{part:identified_technos}
%

\subsection{Gratings and waveguide phased array technologies} % (fold)
\label{part:gratings_waveguide_array}

Grating and waveguides phased array identified technologies are presented in this section, and their characteristics are summarized in Table~\ref{tab:gratings_comparison}.

%%%%%%%%%%%%%%%%%%%%%%%%%%%%%%%%%%%
\begin{table*}[b]
	\caption{Gratings and waveguide phased arrays characteristiscs and merit criteria \label{tab:gratings_comparison}}
	\begin{center}
	\begin{tabular}{llllll}
	\hline \hline
	\bf{Criteria} & \bf{CGS} & \bf{AWG} & \bf{SHD} & \bf{PCSP} \\
	\hline
	Spectral domain [nm]& 400-800 & 1450-1800 & 400-700 & 1550-1600 \\
	Spectral resolution $\SR$ & 150-200 & 2000 & 800 & 500\\
	Type & SM, MM &  SM, M-SM & SM, M-SM & SM\\
	$S\omega$  & 400 &  1 & 1 & 1 \\
	Simultaneous spectra & 1 &  12& 30 & 1\\
	%Reduced volume $m$ & $5\cdot 10^{6}$& $3\cdot 10^{9}$& $6\cdot 10^{8}$ & $8\cdot 10^{9}$ & $3 \cdot 10^{5}$ \\
	%%Operability & $\frownie$ & $\smiley$ & $\smiley$ & $\frownie$ & $\smiley$ \\
	%Efficiency & 0.3 & 0.5 & 0.7 & 0.1 \\
	%Sampling & parallel & parallel & parallel & parallel & parallel\\
	%Polarization & 2 pol &  2 pol & 2 pol  & 2 pol \\
	3D multiplexing & Bad & Possible & Bad & Possible\\
	TRL & 4 & 6  & 4 & 3\\
	\hline
\end{tabular}
\end{center}
\end{table*}
%%%%%%%%%%%%%%%%%%%%%%%%%%%%%%%%%%%

\subsubsection{Semi-integrated Compact Grating Spectrometer (CGS)} % (fold)
\label{part:semi_integrated_compact_grating_spectrometer}

\begin{figure}[t]
  \center
  \includegraphics[height=4.5cm]{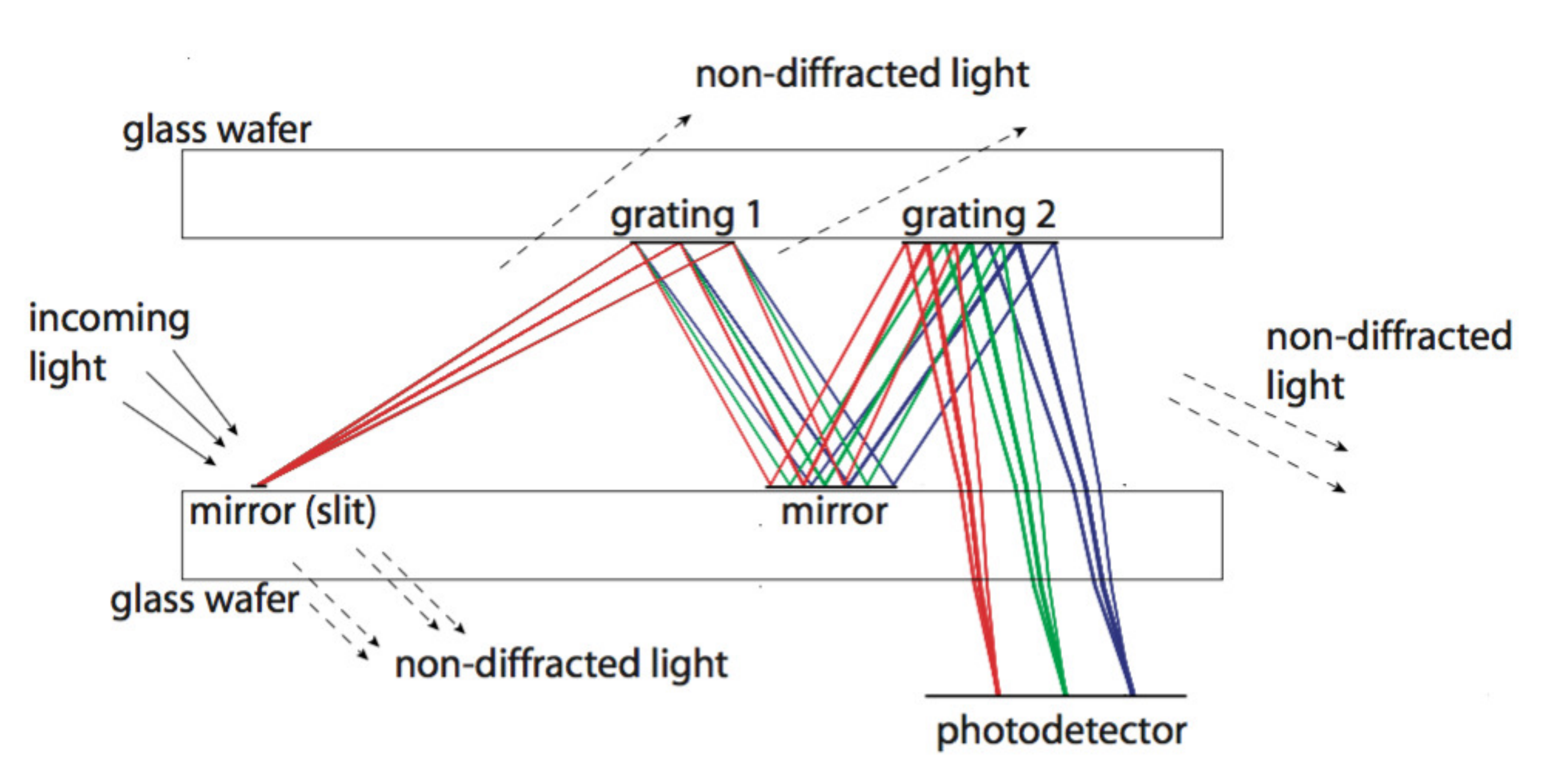} \hfill
  \includegraphics[height=4.5cm]{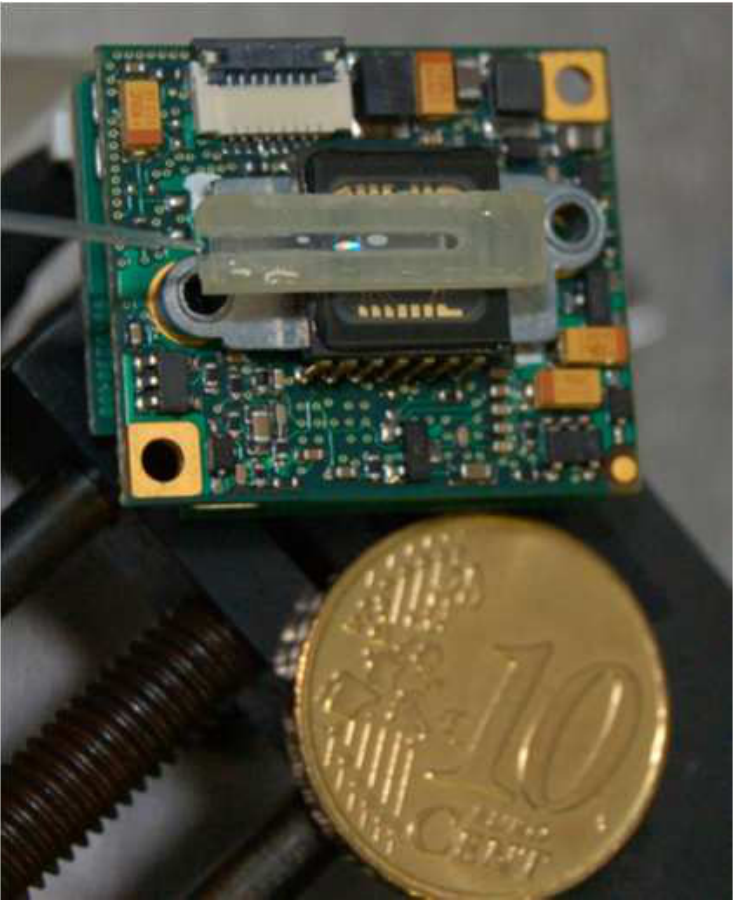}
  \caption{{\it Left:} CGS concept showing the double dispersion stage. The light fed the spectrograph from a fiber in the left. {\it Right:} CGS prototype, compared to a 10 cents coin. The spectrometer is mounted atop the sensor, with grating visible as a tiny ovoidal part.}
  \label{fig:compact-grating-spectro}
\end{figure}

\cite{grabarnik_2007a} introduced a concept of miniaturized spectrometer using two consecutive gratings inserted between two glass plates [Fig.~\ref{fig:compact-grating-spectro}]. An elliptical concave mirror refocuses the beam, setting an image of entrance slit on detector. The beam passes a second time on a grazing (razing grating) that increases the dispersion.  The efficiency also suffers from the grating double stage. In-chip multiplexing is allowed by setting an image of sky on the entrance slit. Since it is a semi-integrated system, the alignment of the slit mirrors and gratings has to be very precise and stable for not loosing light and part of the spectrum. Also, because the grating is a non collimated beam spectral resolution is limited.

% subsubsection semi_integrated_compact_grating_spectrometer (end)

\subsubsection{Arrayed Waveguide Grating (AWG)} % (fold)
\label{part:awg_and_planar_gratings}

\begin{figure}[tbp]
  \center
  \includegraphics[height=5cm]{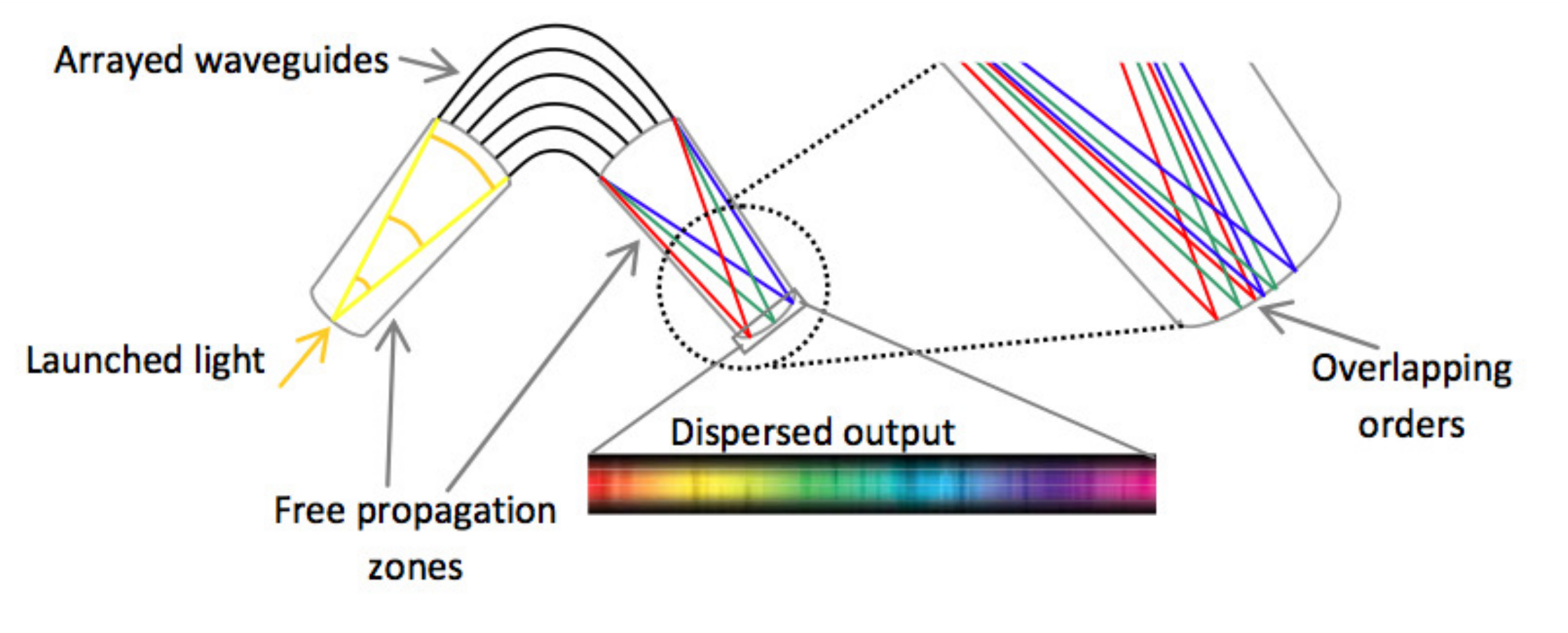}
  \caption{Schematic concept of the AWG \cite{lawrence_2010a}. The light is progressively delayed between the waveguides to reach  high spectral resolution (as would be done by the facets of a blaze grating) before recombining in a free propagation area and forming the spectrum on the detector.}
  \label{fig:awg}
\end{figure}

Fully integrated planar spectrometers have been intensively developed for telecommunications as multiplexers, and are now studied for astronomical applications. \cite{bland_2006a} review these spectrometers for astrophysical applications. We only  present here the Arrayed Waveguide Gratings (AWG) which is the technology that has been the most developed during the last years \cite{cvetojevic_2009a, cvetojevic_2012a} with particular efforts towards high resolutions $\ge 5\times 10^4$. The spectrum is obtained thanks to interference between light traveling through a phased array of waveguides of different lengths [Fig.~\ref{fig:awg}]. A similar concept \cite{bock_2012a} makes use of a waveguide grating instead of a phased array to modulate the phase.
AWG allows multiple SM inputs (up to 12) for in-chip multiplexing or a M-SM use \cite{cvetojevic_2012a}. AWG are also fully integrated spectrographs, so that it is easy to stack several of them in front of a single detector \cite{bland_2010a}.
AWG are fully integrated spectrographs with no moving parts, and are therefore extremely stable.
Spectral resolutions up to 10000 can be achieved, but spectral orders can superimpose if the free spectral range is too small. The bandwidth should be restricted, or a cross dispersion stage should be implemented \cite{lawrence_2010a} at the cost of size \cite{harris_2012b}. The efficiency of the additional off-axis inputs is also lower, reducing the effective spectral resolution by $\sim 20\%$ \cite{cvetojevic_2012a}. Obtaining an extremely uniform index between the waveguides and in the free-propagation area can limit the capacities of mass productions.

% subsubsection awg_and_planar_gratings (end)

\begin{figure}[btp]
  \center
  \includegraphics[height=4.cm]{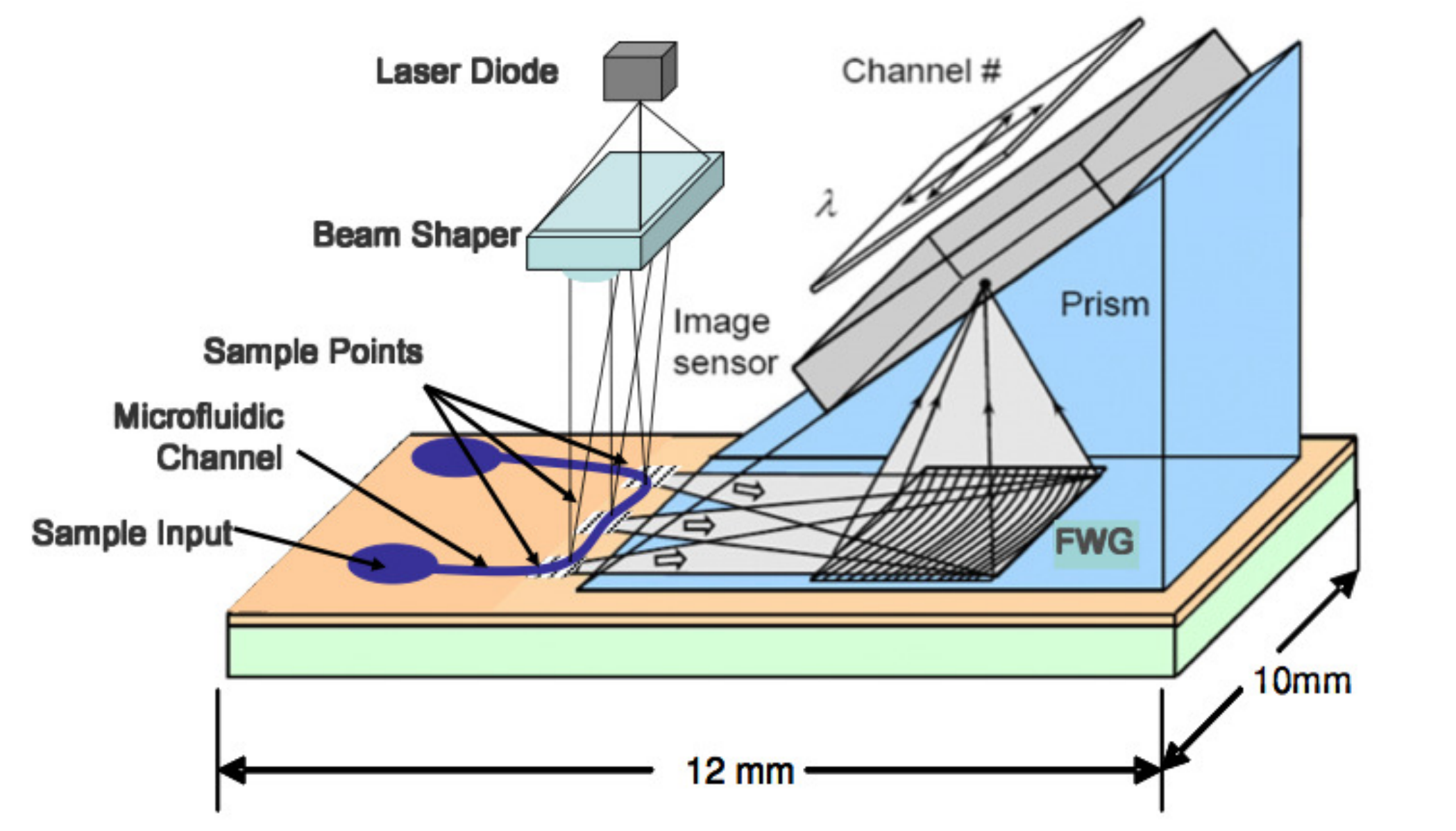}
  \caption{Schematic of the SHD spectrometer \cite{bergner_2012a}, as proposed for micro-fluidic applications: a fluid under test is placed at the entrance of the input waveguides, and is illuminated from the top by a laser diode + beam shaper. For astrophotonics application, the fluid reservoir would be removed to make space for a series of input waveguides coming from the telescope, and laser diode and beam shaper are not present. The light is then introduced on the edge of the planar waveguide along the arrow on the left, and is diffused on top of the grating zone, before interfering on the detector.}
  \label{fig:side-holographic}
\end{figure}
% subsubsection side_holographic_dispersion (end)

\begin{figure}[hbtp]
\center
  \includegraphics[height=4.5cm]{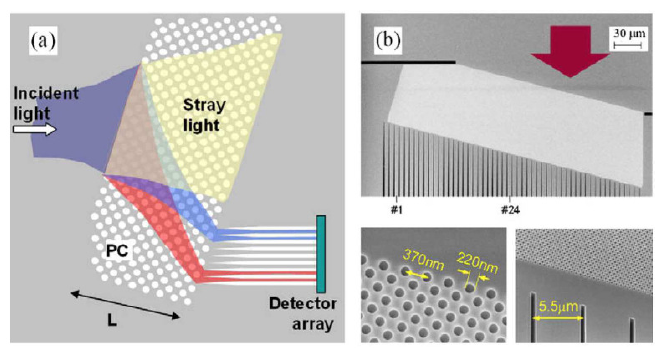}
\caption{(a) Schematic of the operating principle of the fabricated PC spectrometer, which combines three dispersive properties of PCs. (b) Scanning electron microscopy (SEM) image of the PC structure with insets showing details of the interface of the PC and the array of waveguides that sample the optical beam at the output of the PC region.}
%  \caption[crystal-photonic-prim] {{\it Top, left:} Optical microscope photograph of the PCS \cite{lupu_2004a}. We distinguish several inputs on the left, the semi-circular super-prism in the center and the output waveguides on the right, each corresponding to a different spectral channel. {\it Top, right:} Scanning electron microscope on the central photonic crystal area. {\it Bottom:} Operation principle of the negative index super-prism crystal photonics device of \cite{momeni_2009a}.}
  \label{fig:crystal-photonic-prism}
\end{figure}

\subsubsection{Side Holographic Disperser (SHD)} % (fold)
\label{part:side_holographic_dispersion}

Side Holographic Disperser \cite{avrutsky_2006a} couples a grating to a detector without intermediate optics [Fig.~\ref{fig:side-holographic}]. The light is introduced in a planar waveguide, on top of which grating facets are drawn, permitting light to outcouple and interfere on the detector. It is therefore similar to echelle gratings used with an infinite tangent, which allows important spectral resolutions. Multiplexing several SHD is possible but complicated because of the difficulty to align the detector with the focal plan of the spectrograph. SHD can potentially be fed with up to 30 different SM inputs for in-chip multiplexing, but will require cross-dispersion. The focus plan is curved, so that the image sharpness limits the spectral resolution on the edges for broad bandwidth. Using curved detectors \cite{dumas_2012a} could solve this issue and allow a system totally free of optics.

\cite{martin_2015a}  proposes a similar concept in LiNbO3 substrate with a unique single mode waveguide and nano-grooves etched directly on the top.

%\newpage
\subsubsection{Photonic Crystal SuperPrism (PCSP)} % (fold)
\label{part:crystal_photonic_super_prism}

\cite{kosaka_1998a, kosaka_1999a} demonstrated a {\it super-prism phenomenon} in photonic crystals, where deviations angles are several hundred times higher than in conventional prisms, hence delivering spectral resolution of a few hundreds. Different practical implementations have then been proposed. \cite{lupu_2004a} developed a prism in a very tiny volume using Silicon-On-Insulator technology, which has the advantage of allowing multiple inputs, at the cost of a cross dispersion stage and performance depending strongly on the injection angle. \cite{momeni_2009a, gao_2015a} propose concepts take advantage of the negative index phenomena to isolate the dispersed signal from stray light [Fig.~\ref{fig:crystal-photonic-prism}]. The planar nature of such devices allows to stack many of them in a compact way. These are very compact, fully integrated spectrographs that allow easy operations and stability.

%\newpage 
\subsection{Fabry-P\'erots and filters technologies} % (fold)
\label{part:fabry_perots_filters}

Filters and Fabry-P\'erot interferometers identified technologies are presented in the following sections, and their characteristics are summarized in Table~\ref{tab:FP_comparison}.

%%%%%%%%%%%%%%%%%%%%%%%%%%%%%%%%%%%%%%%%%%%%%%%%%%%%%%%%%
\begin{table*}[t]
	\caption{Merit criteria of Fabry-P\'erot, filters \& sorters \label{tab:FP_comparison}}
	\begin{center}
	\begin{tabular}{lllll}
	\hline \hline
	\bf{Criteria} &\bf{PCOS}  & \bf{CCR} & \bf{MEMS} & \bf{PPSI}  \\
	\hline
		Spectral domain [nm] & 400-700 & 400-2000 & 1000-1800 & 400-700 \\
		Spectral resolution $\SR$ & 10 & 10-10$^5$ & 100 &5 \\
		Type & MM & MSM, SM & MM & MM \\
		$S\omega$ & 1 & 1 &100& 10 \\
		Simultaneous spectra & 10 & 1000 & 1000 & $10^{6}$ \\
		%Reduced volume $m$ & & & $3\cdot10^3$ & $1\cdot 10^4$ \\
		%%Operability & & $\smiley$ & $\smiley$ & $\smiley$  \\
		%Efficiency & $\ge 0.7$ &  & 0.5 & 0.1 \\
		%Sampling & parallel & parallel & parallel & parallel \\
		%Polarization & 2 pol  & 1 pol  & 2 pol & 2 pol   \\
		3D multiplexing & Very good & Very good  & Good & Very good \\
		TRL & 4 & 8 & 5 & 4 \\
%		Astrophysical targets & No ! & Compact & Wide field & UBV Photometry\\
		\hline
\end{tabular}
\end{center}
\end{table*}
%%%%%%%%%%%%%%%%%%%%%%%%%%%%%%%%%%%%%%%%%%%%%%%%%%%%%%%%%

\subsubsection{Photonics Crystal Outcoupling Spectrometer (PCOS)} % (fold)
\label{part:photonic_crystal_outcoupling_spectrometer}

\begin{figure}[b]
\begin{centering}
  \includegraphics[height=4cm]{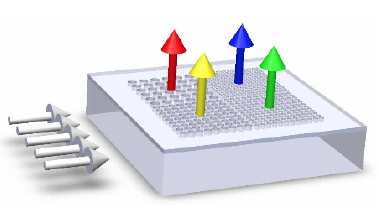} \hspace{1.5cm}
  \includegraphics[height=4cm]{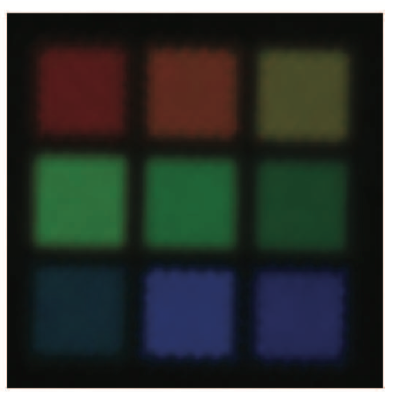}
  \caption[] {{\it Left:} Principle of the PCOS spectrometer. {\it Right:} 3x3 spectrum of an LED obtained by stacking a PCOS on top of a commercial CMOS camera.}
  \label{fig:PCOS}
\end{centering}
\end{figure}

Photonics Crystal Outcoupling Spectrometer \cite{pervez_2010a, gan_2012a} consist in a cascade of photonic crystal patterns drawn on top of a waveguide and able to outcouple its electromagnetic field [Fig.~\ref{fig:PCOS}]. Conversely to \cite{pervez_2010a}, we would imagine a linear arrangement of PCOS channels along a linear MMF or SMF, to not lose photons. Because the photonic crystals structure is drawn on a glass slide, it can be easily stacked on a 2D detector without need for additional optics. With dimensions as small as $30 \times 30 \mu m$ per outcoupling element, they appear compatible with science detectors used in astronomy. They could make the pixel use efficiency of 2D detectors for integral field spectroscopy extremely high, with up to 1 pixel per spectral element.

It is easy to produce once a high quality master is done. If PCOS is used as cross-disperser for an instrument with $\SR \ge$ 1000, multiplexing capability will mostly depend on the previous spectrograph.

\subsubsection{Cascaded Ring Resonators (CRR)} % (fold)
\label{part:filter_ring_resonator}

\begin{figure}[th]
\begin{centering}
  \includegraphics[height=3.5cm]{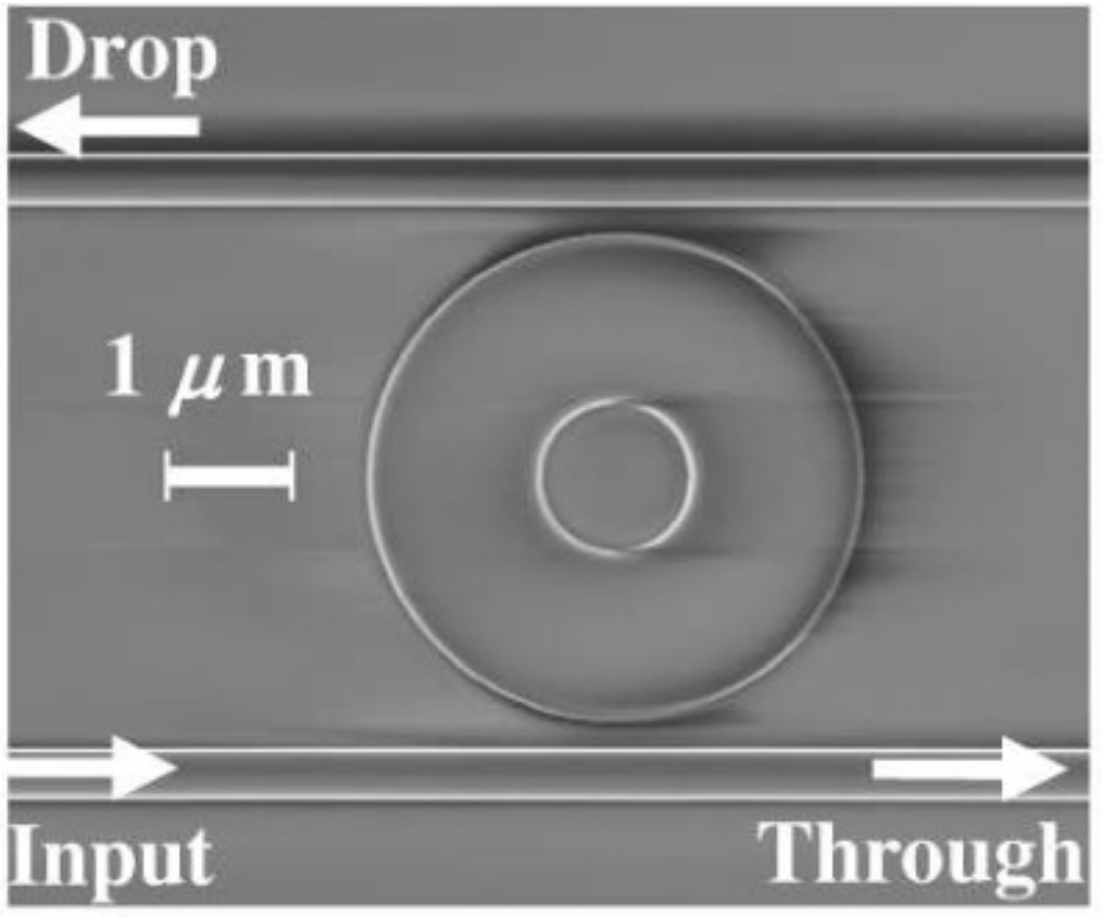} \hfill
  \includegraphics[height=3.5cm]{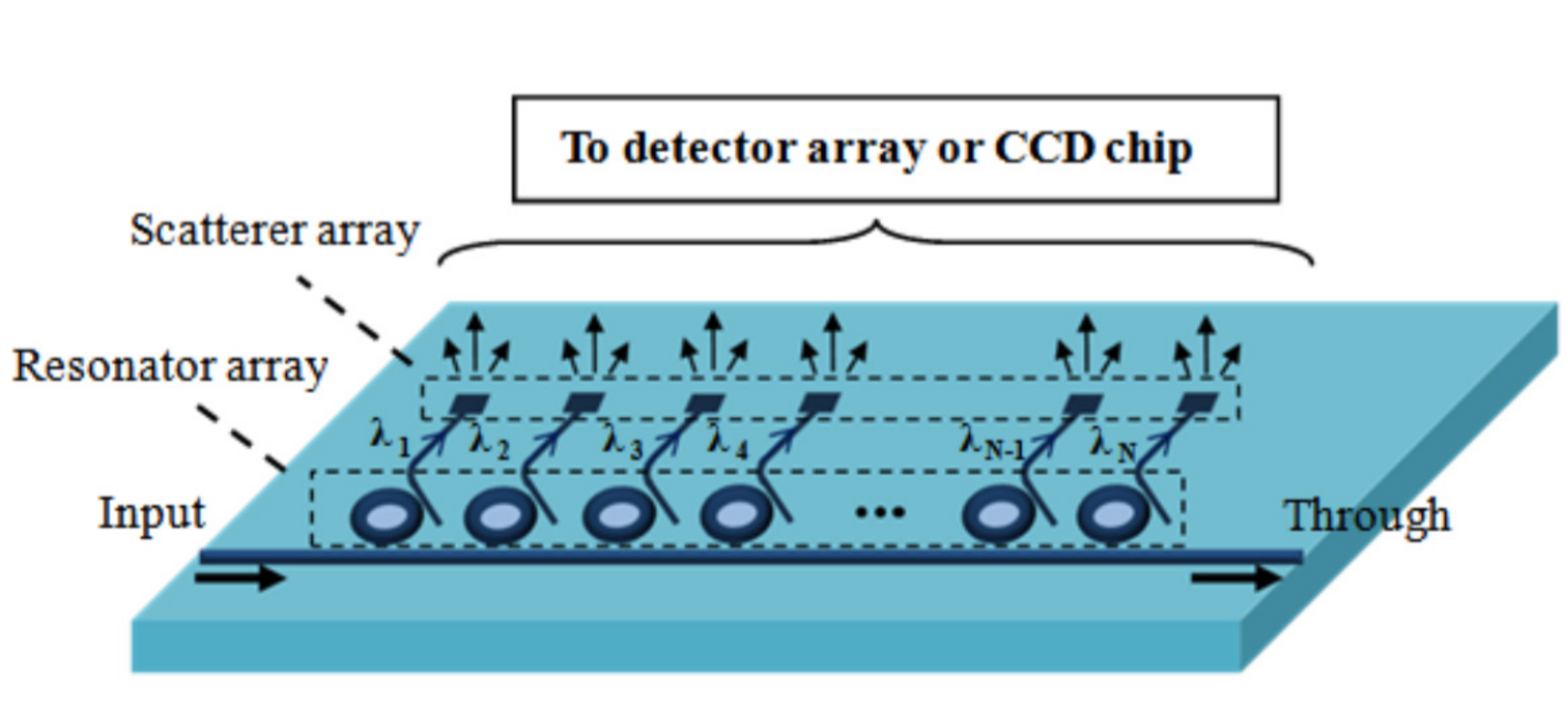}
  \caption[] {{\it Left:} Scanning Electron Microscopy image of a 1.5$\mu m$ radius resonator ring with input and output waveguides \cite{soltani_2010a}. {\it Right:} Concept of cascaded ring resonator \cite{xia_2011a}.}
  \label{fig:CRR}
\end{centering}
\end{figure}

High quality optical microcavities like microrings \cite{bogaerts_2012a} and microdisks \cite{soltani_2010a} can provide ultra-compact high resolution filters, with high-sensitivity, and easy realization [Fig.~\ref{fig:CRR}]. Such rings have been realized with radii down to 1.5$\mu$m  \cite{soltani_2010a}, much smaller than typical pixel sizes used in astronomy, while reaching a spectral resolutions above $10^5$. Large spectral coverage is achieved by cascading such devices, and large free spectral ranges ($\ge$ 50nm) make such devices compatible with energy sensitive detector. We can think of coupling them with Photonic Crystal Outcoupling Spectrometer (Sect.\ref{part:photonic_crystal_outcoupling_spectrometer}) to keep a 2D and highly stable assembly, with the detector directly glued on top of the spectrometer substrate. A fine tuning of the free spectral range to the resolution of the PCOS would allow to limit the number of rings to draw.
Multiplexing is easy via cascading such rings along an input waveguide. Signal is recoupled to a second waveguide on the opposite side, to bring it to the detector [Fig.~\ref{fig:CRR}]. 
The low temperature sensitivity of the order 0.1nm/K and high compactness make them suitable for high precision radial velocities (1m/s corresponding to 50mK stability) \cite{xia_2011a}.

%\vfill

\begin{figure}[h]
  \centering
  \includegraphics[height=3.5cm]{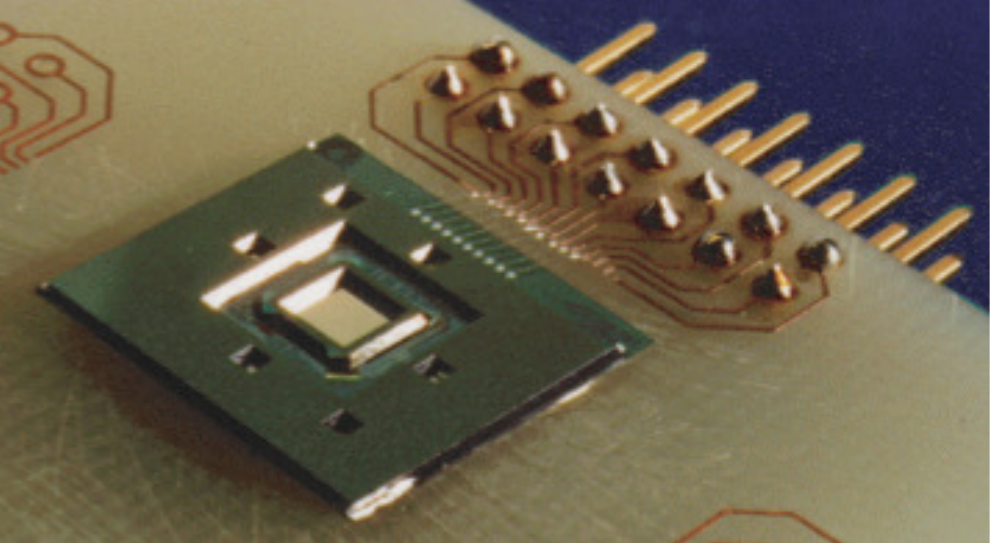}\hspace{1.5cm}
  \includegraphics[height=3.5cm]{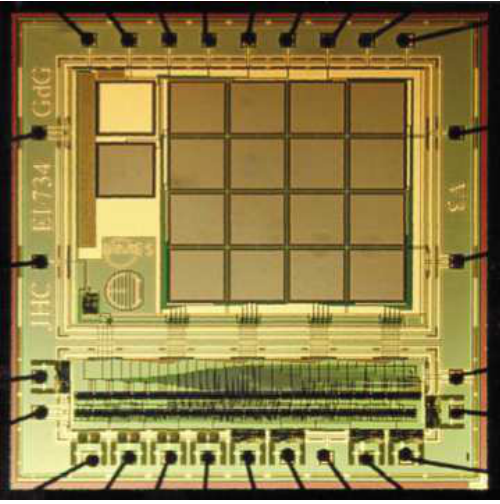}
  \caption{MEMS-based optical mini spectrometers, with on top the schematics of the tunable Fabry-P\'erot cavity, and on bottom the actual integrated spectrograph \cite{wolffenbuttel_2005a}.
  \label{fig:mems_tunable} Photographs of an on-ship  array of MEMS-based Fabry-P\'erots \cite{wolffenbuttel_2005a}.}
  \label{fig:mems_array}
\end{figure}

%\vfill  

\begin{figure}[h]
  \center
  \includegraphics[height=3.8cm]{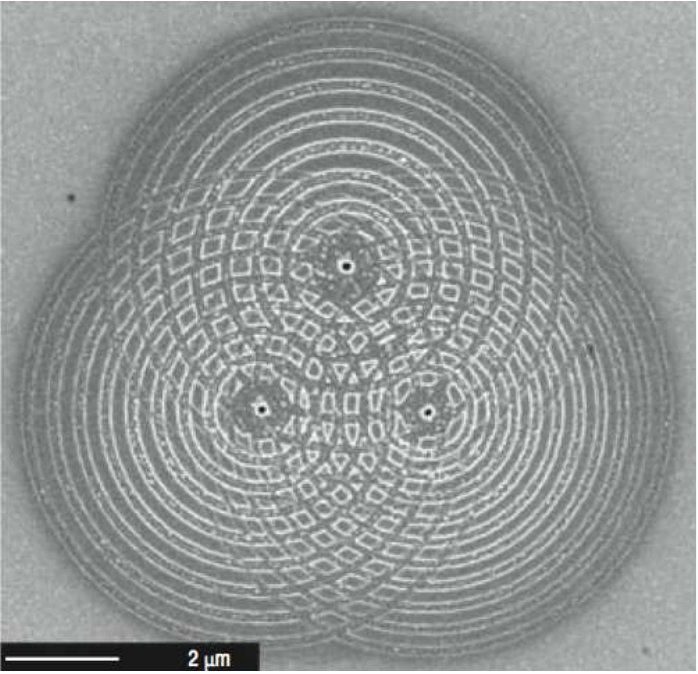} \hspace{2cm}
  \includegraphics[height=4.5cm]{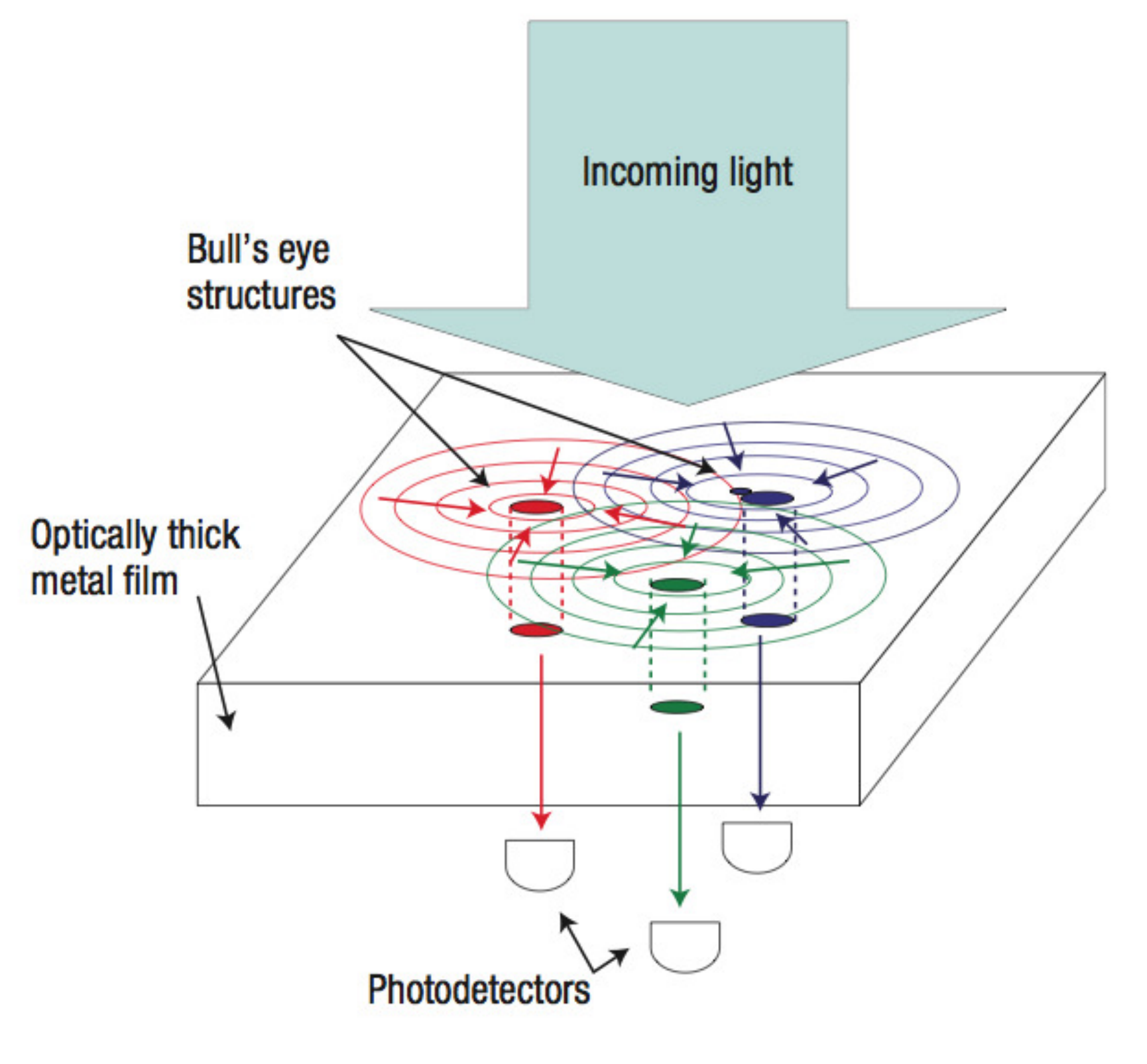}
  \caption[plasmonic-sorter] {{\it Left:} Series of three antennas with grooves periods of 530 nm (right), 630 nm (left) and 730 nm (top) to sort the corresponding wavelengths. {\it Right:} Schematics of the 3-antenna PPSI plus the detector behind the metal film \cite{laux_2008a}.}
  \label{fig:plasmonic-sorter}
\end{figure}

\newpage

\subsubsection{MEMS-Based optical mini spectrometers (MEMS)} % (fold)
\label{part:mems_based_optical_mini_and_micro_spectrometers}

\cite{wolffenbuttel_2005a} describe semi-integrated MEMS-based Fabry-P\'erot spectrometers. It allows to tune the cavity to the desired wavelength, with a limited spectral range [Fig.~\ref{fig:mems_tunable}]. A second concept uses 16 Fabry-P\'erot on the same chip, with fixed cavity spacing. The spectral resolution can then reach a few hundreds, but the small number of FP limits the spectral range. Multiplexing is possible by using fiber bundles to feed each chip. An array of MEMS-based FP can be conveniently set directly at the focus of the telescope with a large field-of-view for moderate spectral resolutions, but the spatial filling factor could be low because of the surface of the chip with respect to the filter area.

% subsubsection mems_based_optical_mini_and_micro_spectrometers (end)

\subsubsection{Plasmonic Photon Sorters Imaging  (PPSI)} % (fold)
\label{part:plasmonic_sorters_imaging}

\cite{laux_2008a} describe a novel way to detect and sort photons at the same time. A series of periodic grooves of wavelength-size is drawn at a metal surface and acts like an antenna for the incoming light. By intricating several of these antennas [Fig.~\ref{fig:plasmonic-sorter}] with a different grooves periods (each being dedicated to a single wavelength), photons approaching the metal surface are sorted and directly go to the proper antenna. Conversely to others filters, no photons is reflected so that the SNR of such a device is close from a gratings. The relatively small size of antennas could allow a reduction in detectors size. It is also possible to draw antennas that are polarization sensitive.
A series of such antennas in front of a detector array should allow direct spectro-imaging without optics (i.e.~fiber bundles or slicers).
Such a filter can be put directly in front of a detector and does not require optics, it should be extremely stable.
The efficiency and spectral resolution of each antenna depending on the quality factor of the grooves, a NIR concept should be of better quality than a visible one. But it has not been demonstrated that high spectral resolution can be reached (influence area of antennas, pixel size, etc.).

\subsection{Fourier Transform Spectrometers technologies} % (fold)
\label{part:fourier_transform_spectrometers}

%%%%%%%%%%%%%%%%%%%%%%%%%%%%%%%%%%%%%%%%%%%%%%%%%%%
\begin{table*}[b]
	\caption[Comparison FTS]{Merit criteria comparison of FTS \label{tab:FTS-compare}}
	\begin{center}
	\begin{tabular}{lllllll}
	\hline \hline
		\bf{Criteria} & \bf{$\mu$SPOC}  & \bf{SPOC-HR} & \bf{SWIFTS} & \bf{LLIFTS}  & \bf{AMZI}\\
		                   &               (slit)       &                         &                    &                    &               \\
		\hline
		Spectral domain [nm]& 3000-5000 &  500-1100 & 400-1000 & 1200-1900 & 1500-1600 \\
		Spectral resolution $\SR$ & 200 & 4000 & 100000 &250 & 4000  \\
		Type & MM & MM & SM, M-SM & SM & M-SM \\
		$S\omega$ & 2000 & > 10000  & 1-256 & 1 & 50-200 \\
		Simultaneous spectra& $>$ 300 & 1 & 1& 1 & 1 \\
		%Reduced volume $m$ & $2\cdot 10^{3}$& $5\cdot 10^{5}$& $3\cdot 10^{5}$ & $3\cdot 10^{7}$ &$3\cdot 10^8$\\
		%%Operability & $\smiley$ & $\smiley$ & $\smiley$  & $\smiley$ & $\frownie$\\
		%Efficiency & 0.2 & 0.1 & 0.3 & 0.7 & 0.5 \\
		%Sampling & Static & Static & Static & Static & Static \\
		%Polarization & 2 pol & 1 pol & 1 pol & 1 pol &  1 pol \\
		3D multiplexing & No & No & Possible & Bad & Possible \\
		TRL & 5 & 5 & 9 & 4 & 4 \\
%		Astrophysical targets & Compact & Compact & Compact & Compact & Compact\\
%                                  & Faint, wide &            &                  &               & Faint\\
		\hline
\end{tabular}
\end{center}
\end{table*}
%%%%%%%%%%%%%%%%%%%%%%%%%%%%%%%%%%%%%%%%%%%%%%%%%%%

FTS identified technologies are presented in the following sections, and their characteristics are summarized in Table~\ref{tab:FTS-compare}.

%%\newpage
\subsubsection{SPectrometer On Chip (SPOC)}
\label{part:microspoc_fourier_transform_spectrometers}

\begin{figure}[b]
  \center
  \includegraphics[height=4.cm]{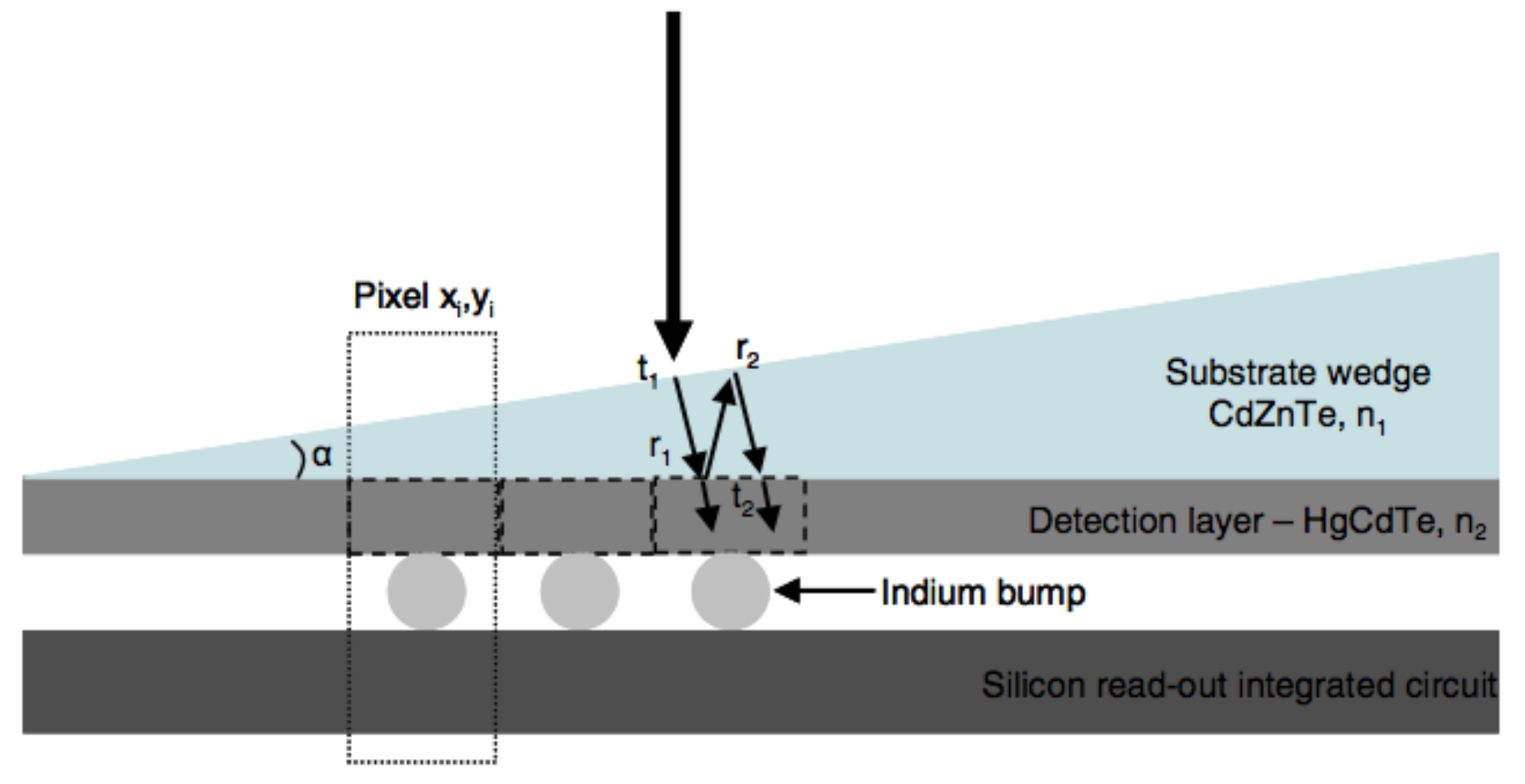}\\ \vspace{2em}
 \includegraphics[height=5cm]{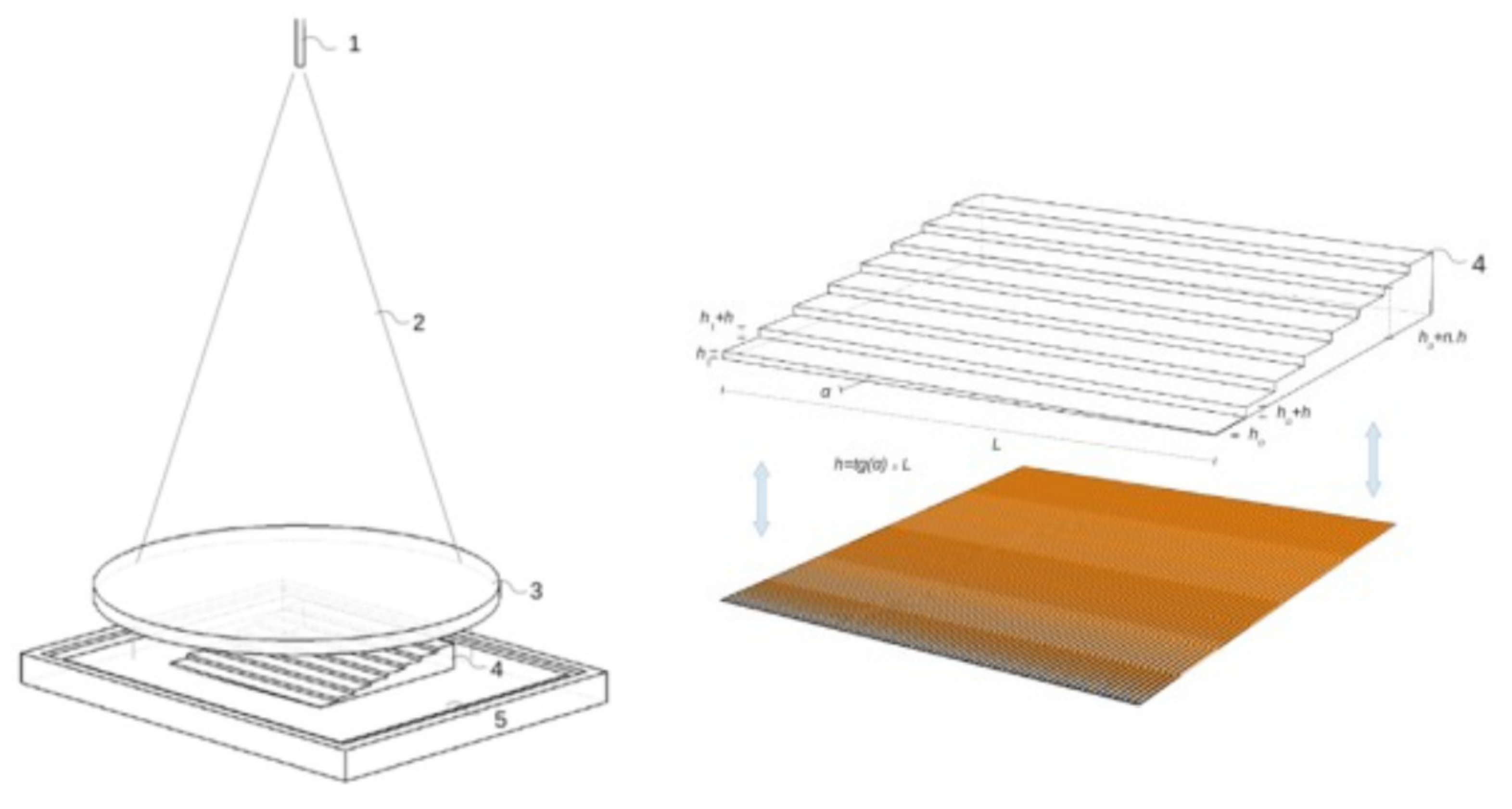}

  \caption{{\it Top:} Schematic of the $\mu$SPOC concept \cite{gillard_2012a}. The delay is created by multiple reflections in the wedged substrate. {\it Bottom:} SPOC-LA concept using a stepped wedge to increase spectral resolution and/or spectral coverage \cite{lecoarer_2014a}.}
  \label{fig:microspoc}
\end{figure}

Different versions of the SPectrometer On Chip (SPOC) have been developed. The initial concept \cite{guerineau_2005a, rommeluere_2007a} is a focal plane, static FTS, where the OPD modulation is generated with a wedged substrate atop the detector [Fig.~\ref{fig:microspoc}]. The detector is immersed in a medium of optical index $n\sim2.7$, increasing the field of view by a factor $\sim 8$ according to the classical case. This concept allows different spectro-imaging modes depending on the optical layout:
\begin{enumerate}
\item In a focal plane, it works like an imaging system, presenting interference fringes all over the field. By scanning the field in the direction of the fringes, we modulate the OPD for each point of the image. $\mu$SPOC is then a spectro-imager. However, some points of the field are not used during this procedure and the fringe modulation is not static.
\item $\mu$SPOC can be used like a slit spectrometer by using a cylindrical lens, and scanning the field in the transverse direction. All the pixels are useful and the fringe sampling is static (conversely to case 1).
\item In a pupil plane, $\mu$SPOC observes an extended source at finite distance (the image of the source by the telescope), so that the spatial information over the whole field is averaged on each pixel. This mode is useful for faint and/or extended sources thanks to the important optical \'etendue.
\end{enumerate}
This is a very compact, versatile, and fully integrated concept that can be directly at the telescope focus. The spectral resolution is limited to a few thousands by the maximum OPD at the final edge of the detector. A high spectral resolution version has more recently been developed, called SWIFTS-LA (SWIFTS Large Aperture) \cite{lecoarer_2014a} or SPOC-HR \cite{diard_2016a}. In this configuration the 2$^{nd}$ dimension of the 2D detector is also used for optical path modulation, and spectro-imaging capability is lost. Since SPOCs are fixed delay FTS, inhomogeneities of the substrate and surface errors can be calibrated and compensated in an optimized non-linear Fourier Transform.

\subsubsection{Stationary Wave Integrated Fourier Transform Spectrometer (SWIFTS)}
\label{part:swifts_fourier_transform_spectrometers}

\begin{figure}[t]
  \center
  \includegraphics[height=3.5cm]{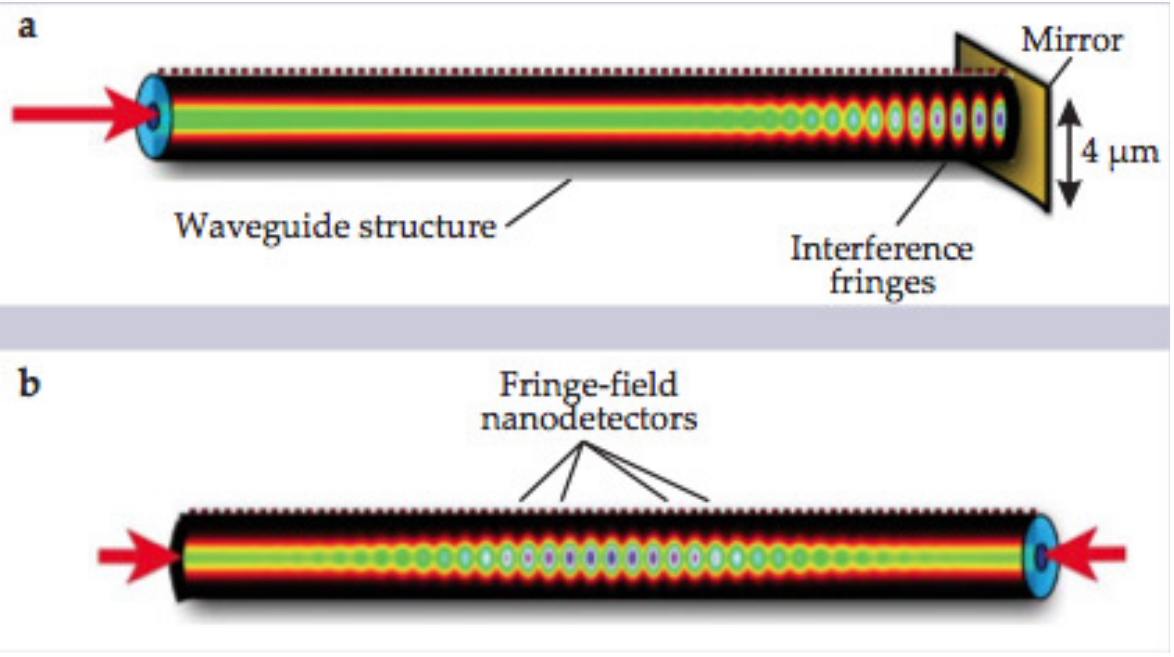}  \hfill
  \includegraphics[height=3.5cm]{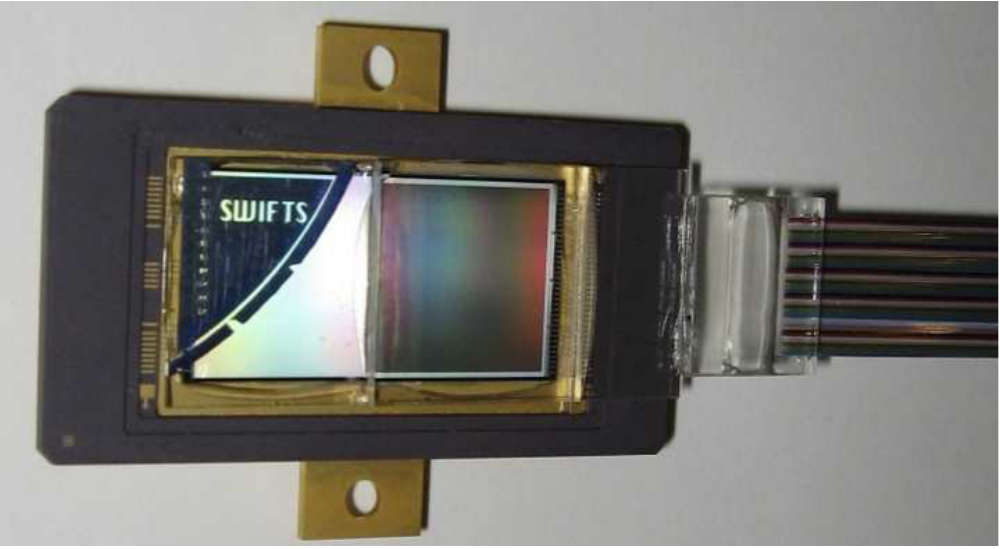}
	\caption[FTS SWIFTS] {{\it Top:} SWIFTS principle in {\it a)} Lippmann and {\it b)} counter-propagative configurations \cite{bland_2012a}. {\it Bottom:} Photograph of a Lippmann SWIFTS prototype containing 64 waveguides, each sampling a different part of the fringe pattern. It can cover the full visible band with a spectral resolution of 100000.}
  \label{fig:SWIFTS}
\end{figure}

{\it Stationary Wave Integrated Fourier Transform Spectrometer} (SWIFTS) \cite{lecoarer_2007a} is a fully integrated FTS where light is diffused to the detector thanks to small gold dots placed along the waveguides with a pitch equal to the CCD pixel one. It exists two different configurations:
\begin{enumerate}
\item Lippmann mode [Fig.~\ref{fig:SWIFTS}, left, (a)] where the forward propagating wave is rejected on a mirror, leading to a stationary wave. The zero delay is then on the mirror.
\item Gabor mode [Fig.~\ref{fig:SWIFTS}, left, (b)], or counter-propagative mode, where light is split in two beams before being introduced into opposite ends of the waveguide.
\end{enumerate}
As fringes are under-sampled in a single waveguide, matrix versions of SWIFTS consist in a collection of parallel surface waveguides glued on a 2D detector, each waveguide sampling a different part of the fringe pattern. They then allow to extent the bandwidth and the optical \'etendue of the spectrometer. A prototype has been designed to reach spectral resolutions higher than $10^5$ over the full visible bandwidth (from 400 to 1000 nm; Fig.~\ref{fig:SWIFTS}, bottom). SWIFTS has now reached the commercial state.

A low noise and very fast version using Superconducting Nanowire Single Photon Detectors has also been developed \cite{cavalier_2011a}.
Spectro-imaging at very high spectral resolution is easily obtained by stacking SWIFTS in a compact fashion and feeding them with fiber bundles sampling the focus plan of the telescope.
SWIFTS is directly coupled to fibers and the detector is directly glued above the waveguides, insuring an extremely stable instrument.
Like for other FTS, SWIFTS resolution is mainly limited by the precision of the fringe sampling (i.e.~the gold dot position precision), and of the index inhomogeneities and dispersion of the waveguides. This can however be calibrated and compensated with an adapted Fourier Transform.

\begin{figure}[t]
  \center
  \includegraphics[height=3.5cm]{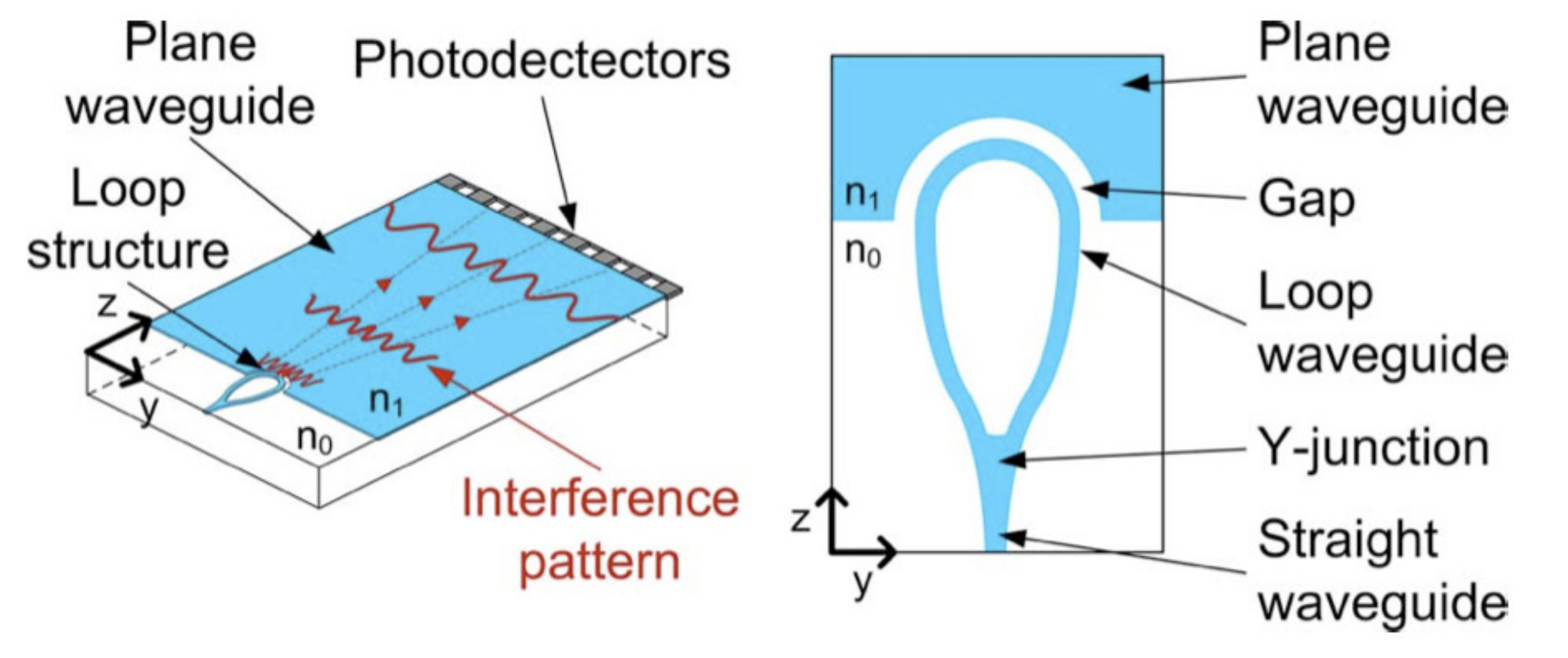} \\ \vspace{2em}
  \includegraphics[height=4cm]{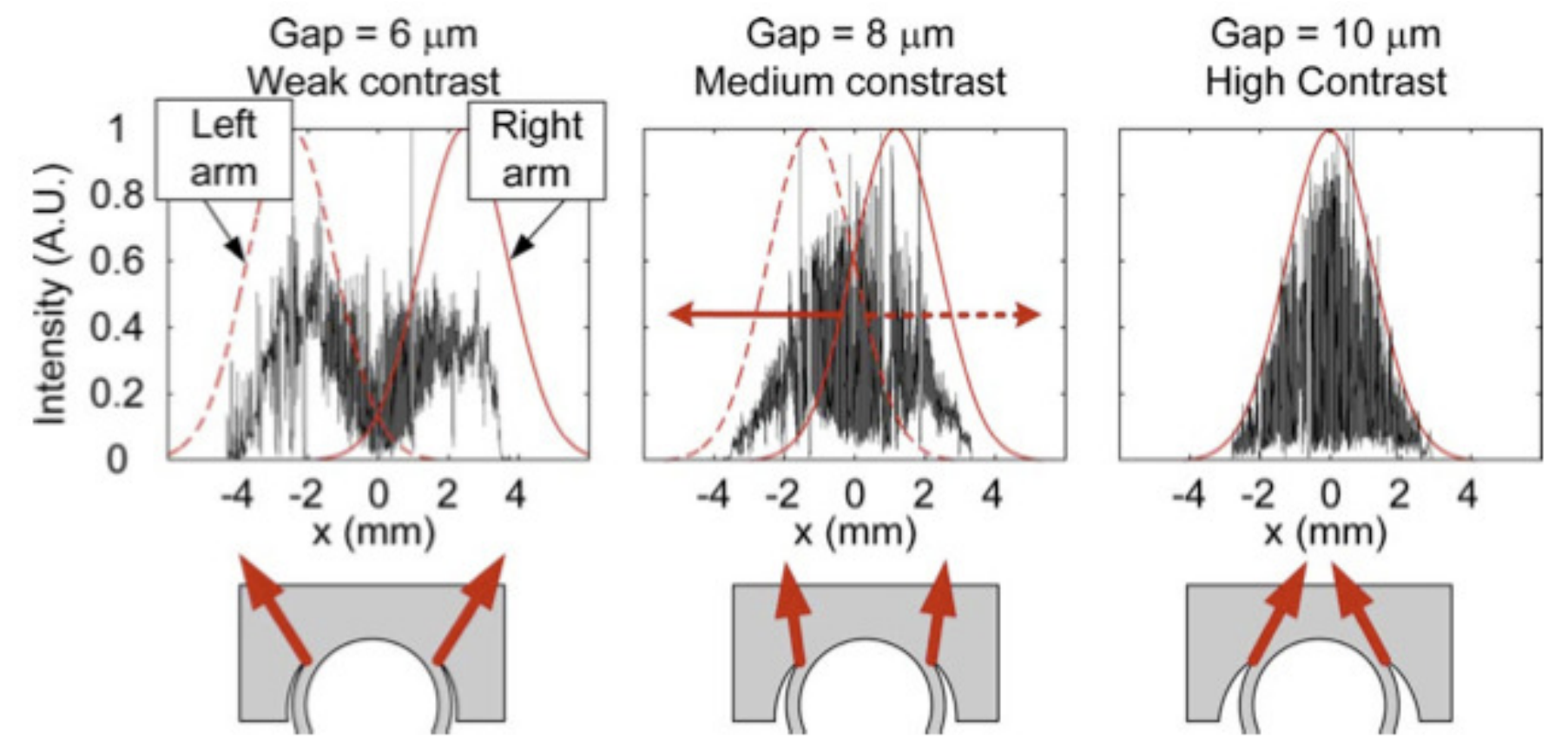}
  \caption[LLIPS] {{\it Top:} Principle of the LLIFTS spectrometer. The light is split in two beams in a loop waveguide. It then progressively leaks from the loop, and interfere in the plane waveguide, at the end of which is the detector \cite{martinB_2009b}. {\it Bottom:} Influence of the loop design on the fringe quality \cite{martinB_2009a}.}
  \label{fig:leaky-loop}
\end{figure}

%%\newpage
\subsubsection{Leaky Loop Integrated Fourier Transform Spectrometer (LLIFTS)}
\label{part:llifts_fourier_transform_spectrometers}

LLIFTS \cite{martinB_2009b, martinB_2009a} consists of a bend waveguide (a loop), from which the light leaks to induce an interference pattern outside [Fig.~\ref{fig:leaky-loop}]. The loop radius and the distance to the detector can be chosen to enlarge the fringe pattern and allow high spectral resolution and/or bandwidth [Fig.~\ref{fig:leaky-loop}, bottom]. The technological locker is a novel way to enlarge the wavefront using the leak in a curved waveguide tangent to a planar waveguide. The gap between two guides permits one to adjust the distribution of light on the edge of component. The distribution of light of the overlapping beams allows a distribution of energy in envelop interferogram giving an apodization optimizing energy. For interferometry, the efficient taperisation permits to extend the concept to multi-telescope Fizeau interferometer as it is proposed \cite{kern_2009a}.
The compact size of this concept allows to stack several of these spectrographs and to put them in front of the detector. Each spectrograph can be directly coupled to a fiber bundle.
The detector is directly set on edge of the planar waveguide and the concept is fully integrated.
The interference of two spherical wavefronts projected on a line detector gives a non linear interferogram, so that a non linear Fourier Transform is required to compute the spectra. The homogeneity of the interference zone has to be excellent.

\subsubsection{Arrayed Mach-Zehnder Interferometers (AMZI)}
\label{part:AMZI}

\begin{figure}[t]
  \centering
  \includegraphics[height=4.5cm]{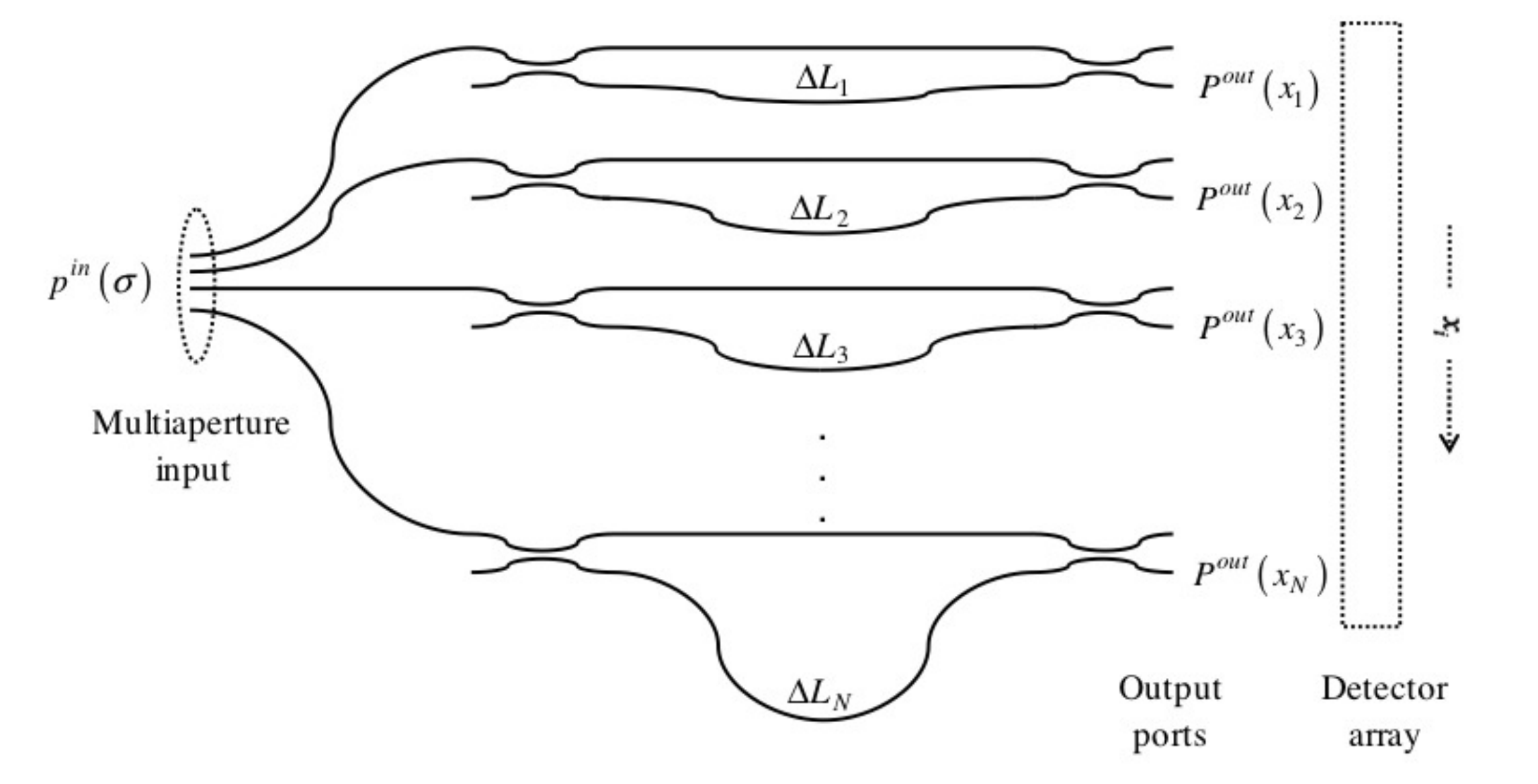} \hfill
  \includegraphics[height=3.5cm]{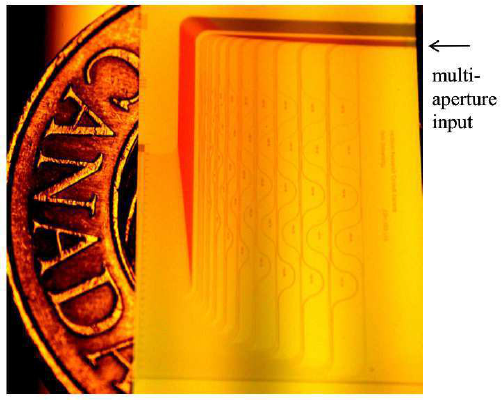}
  \caption[Fabricated AMZI] {{\it Left:} Schematic of the Arrayed MZI spectrometer principle. Each MZI corresponds to a different OPD \cite{florjanczyk_2007a}. {\it Right:} Fabricated silicon chip of AMZI, with 100 s-shaped waveguides \cite{florjanczyk_2009a}.}
  \label{fig:AMZI}
\end{figure}

\cite{florjanczyk_2007a} presents an integrated FTS constituted of an Array of Mach-Zehnder Interferometers (AMZI) [Fig.~\ref{fig:AMZI}] with increasing optical paths. The spectral resolution and the bandwidth depend on the OPD of the most unbalanced MZI, and on the OPD from one MZI to the other respectively. The important advantage of this concept is that each Mach-Zehnder has a final coupler, providing two outputs in phase opposition. Adding them, the instantaneous photometry on each interferometer is measured, and we can compensate, i.e., a non-uniform and varying illumination of the input waveguides. It also means that the number of MZI is equal to half the number of spectral channels. Another interesting property is that each MZI contributes to increase the optical \'etendue of the system. Spectro-imaging is easily obtained by stacking AMZIs in a compact fashion and feeding them with fiber bundles sampling the focus plan of the telescope. The detector is put directly at the component edge and should be very stable. A first version designed with 50 MZIs reaches a spectral resolution $\SR$ = 3600 over 11 nm around 1.55 $\mu$m \cite{florjanczyk_2009a} [Fig.~\ref{fig:AMZI}]. Extending the concept to a bandwidth of 500 nm and a resolution $\dlambda$ = 0.1nm ($\SR$ = $10000$, $\lambda$ = 1 $\mu$m) would require 8000 MZIs, leading to a far more complex, longer (i.e.~less transmissive) and voluminous integrated chip. A more compact version makes use of spiral waveguides \cite{velasco_2013a, velasco_2013b}.

\section{Performance comparison} % (fold)
\label{part:signal_to_noise}

Now that we have introduced the different families of integrated spectrographs and a few actual technologies, we compare the performance of the different concepts depending on the band (MIR, IR, Visible), on the source extent and on the observing conditions.

\subsection{Signal-to-noise ratio} \label{part:SNR_general_expression}

\subsubsection{General expression}
The general expression of the signal-to-noise ratio for spectrographs in a given spectral channel is:
\begin{equation}
\frac{\mbox{S}}{N} = \frac{S_\lambda}{\sqrt{\sigma_{det }^{2} + \sigma_{phot}^{2} + \sigma_{resp}^{2}}},
\end{equation}
where:
\begin{itemize}
\item $S_\lambda$ is the measured signal on a given spectral channel, which expresses as $	S_\lambda$ = $\rho  \cdot  T \cdot  F_{\lambda }\cdot  t$, with $\mathbf{\rho}$ the coupling efficiency in the case of fibered/integrated instruments (Section~\ref{part:coupling_efficiency}). $\mathbf{F_{\lambda}}$ denotes the spectral flux density of the target in one spectral channel. $\mathbf{T}$ is the optical throughput of the instrument, including the detector quantum efficiency $\eta$. It also includes fringe contrast in the case of an FTS (due e.g.~to beams imbalance). $\mathbf{t}$ is the integration time.
\item $\mathbf{\sigma_{det}}$ is the detector noise linked to detector read-out-noise (RON).
\item $\mathbf{\sigma_{phot}}$ is the photon noise, including the source and the background. Table~\ref{tab:background} gives typical values of the sky brightness in optical bands. The sky background increasing in proportion of the observed optical \'etendue, AO are very useful since they concentrate the source energy on the smallest possible area of the sky, then minimizing the background noise.
\item $\mathbf{\sigma_{resp}}$ is the standard deviation of the pixel response fluctuations, or flat-field noise. It can be considered as an additional noise that applies to the signal plus background as a small fraction of it. The noise is close to percent and is negligible in front of photon or detector noise. It is therefore ignored in the following.
\end{itemize}
We describe these noises for the different type of instruments, and remind the identified technologies for each type in Table~\ref{tab:summary_techno}.

%%%%%%%%%%%%%%%%%%%%%%%%%%%%%%%%%%%%%%%%%%%%%%%%%%%%%%%%
\begin{table*}[t]
\centering
\caption[Expected performance of AO on ELTs.]{Expected performance of MCAO/LTAO systems on the 42-m E-ELT (old design) in standard conditions (seeing = 0.8'') \cite{diolati_2010b, fusco_2010c}. Bottom half presents the expected normalized optical \'etendue in different cases. The spatial resolution of 100 mas is a typical E-ELT scientific requirement when diffraction limited images are not required. EE stands for Encircled Energy. \label{tab:strehl_ELT}}
\begin{tabular}{lrrrrrrr}
\hline \hline
 Band & V & I & J & H & K & M & N\\
$\lambda$ [nm] & 500 & 900 & 1250 & 1650 & 2200 & 4800 & 10500 \\
\hline
{\bf Strehl and angles on sky} \\
$\St$ [\%] & 0.5 & 5.5 & 18 & 35 & 50 & 90 & 97 \\
FWHM [mas] & 9 & 8 & 9 & 10 & 12 & 24 & 49 \\
Width for 50\% EE [mas] & $> 100$ & $> 100$ & $100$ & $\sim 30$ & $20$ & $\sim30$ & 49 \\
Width for 85\% EE [mas] & \multicolumn{7}{c}{$> 100$} \\
\hline
\multicolumn{8}{l}{\bf Normalised optical \'etendue S$\omega$} \\
Point + AO ($> 50\% EE$) & $> 1500$ & $> 500$ & 250    & 12  &      3  & 2 & 1 \\
Point + AO ($> 85\% EE$) & $\gg 1500$ & $> 500$ & $>250$    & $>150$  &  $> 80$  & $>16$ & $> 4$ \\
Resolved ($\theta$ =50 mas) & 375 & 125 & 62 & 38 & 20 & 4 & 1 \\
Resolved ($\theta$ =100 mas) & 1500 & 500 & 250 & 150 & 78 & 16 & 4 \\
\hline
\end{tabular}
\end{table*}
%%%%%%%%%%%%%%%%%%%%%%%%%%%%%%%%%%%%%%%%%%%%%%%%%%%%%%%%

\paragraph{\bf Grating, phased array, and cascaded Fabry P\'erot}

In gratings and phased array dispersers, a pixel is equivalent to a spectral channel. Cascaded Fabry P\'erot like Ring Resonators provide equivalent performance because no light is rejected, conversely to a single FP cavity system. They are only sensitive to the light from the source $F_\lambda$ and the background $B_\lambda$ at the given wavelength of the pixel.
\begin{eqnarray*}
%begin{center}
&S_\lambda &= \rho  \cdot  T \cdot  F_{\lambda }\cdot  t\\
&\sigma_{det}^2 &= N_{pix}  RON^2\\
&\sigma_{phot}^2 &= \rho  T t (F_\lambda + 2 B_\lambda)
%end{center}
\end{eqnarray*}
The factor 2 in the photon noise is due to the background which is measured with the source, and subtracted after measurement on sky only.

\paragraph{\bf Fourier Transform Spectrometer}

Each pixel samples part of a fringe which results from the interference of the light from the {\it whole} spectral band of interest, and suffers then from an increased photon noise. The FT then transport the noise accumulated by all the pixel to each spectral channel.
\begin{eqnarray*}
&	S_\lambda &= \rho  \cdot  T \cdot  F_{\lambda }\cdot  t\\
&\sigma_{det}^2 &= N_\lambda  RON^2\\
&\sigma_{phot}^2 &=  N_\lambda\rho  T t (F_{\lambda } + 2B_\lambda)
\end{eqnarray*}

\paragraph{\bf Single Fabry-P\'erot cavities}

In a classical Fabry P\'erot made of a single (scanning) cavity, only one spectral channel can be measured at a time. Hence, to scan a given spectral band, only $1/N_\lambda$ of the whole integration time can be dedicated to each spectral channel.
%
%%%%%%%%%%%%%%%%%%%%%%%%%%%%%%%%%%%%%%%
\begin{eqnarray*}
&S_\lambda &= \rho  \cdot  T \cdot  F_{\lambda }\cdot  t/N_\lambda\\
&\sigma_{det}^2 &= RON^2\\
&\sigma_{phot}^2 &= \frac{1}{N_\lambda} \rho  T (F_{\lambda }+2 B_\lambda)  t
\end{eqnarray*}
%%%%%%%%%%%%%%%%%%%%%%%%%%%%%%%%%%%%%%%

Although a scanning FP plate only cover 1 spectral channel at a time, this is a very simple system to perform integral field spectroscopy. In the framework of astrophotonics it also benefits from an \'etendue advantage since they are bulk systems, hence requiring a much lower number of pixels than M-SM devices, and spectroscopy of much fainter objects as long as large format detectors will not be with RON below $10^{-2}$-$10^{-3} e^-$/pixel. 

%%%%%%%%%%%%%%%%%%%%%%%%%%%%%%%%%%%%%%%%%%%%%%%%%%%%%%%%
\begin{table}[t]
\centering
\caption{Summary of spectrograph types and related identified technologies}\label{tab:summary_techno}
\begin{tabular}{p{4cm}p{7cm}}
\hline \hline
Grating and phased arrays &  CGS, AWG, SHD, PCSP \\
Single Fabry-P\'erot & MEMS \\
Cascaded Fabry P\'erot & PCOS, CRR, PPSI \\
FTS & $\mu$SPOC, SPOC-HR, SWIFTS, LLIFTS, AMZI\\
\hline
\end{tabular}
\end{table}
%%%%%%%%%%%%%%%%%%%%%%%%%%%%%%%%%%%%%%%%%%%%%%%%%%%%%%%%

\subsubsection{Coupling efficiency into SM waveguides}\label{part:coupling_efficiency}

The coupling efficiency $\rho$ into SM waveguides highly depends on the atmospheric conditions and/or on the source type and strongly impact the performance of SM instruments. We give the expression of $\rho$ in case of extended sources.

\paragraph{\bf Unresolved source in turbulent conditions}
\label{part:point_source}

Unresolved sources are sources smaller than the limit of diffraction of the telescope. As soon as a telescope is bigger than the Fried parameter $r_0$ (typically 10 cm in visible and up to $\sim$ 1 m in IR), reaching its diffraction limit requires an Adaptive Optics (AO).  If not at the diffraction limit, the coupling efficiency $\rho$ into a SM waveguide drops quickly \cite{coudeduforesto_2000}:
\begin{equation} \label{eq:rho_strehl}
\rho = 0.8 \;\St,
\end{equation}
\noindent where $\rho_0$ = 0.8 is the maximum coupling efficiency that can be achieved with a circular, unobstructed telescope \cite{ruilier_1998a}, and $\St$ is the Strehl ratio, which can be written as \cite{fusco_2004}:
\begin{equation}
	\St = E_c + (1-E_c) \left(\frac{r_{0}}{D}\right)^2,
\end{equation}
\noindent where $E_c$ = $\exp(-\sigma_{res}^{noll2})$ is the coherent energy and $\sigma_{res}^{noll}$ is the residual phase disturbance of the wavefront after a partial correction by the AO \cite{noll_1976}. For an uncorrected wavefront, the \'etendue of the speckle pattern is $S\omega \sim (D/r_0)^2$. 

Expected AO performance for the E-ELT \cite{diolati_2010b, fusco_2010c} are summarized in Table~\ref{tab:strehl_ELT}. It gives in particular the 50\% Encircled Energy diameter for a point source, with \'etendue exceeding 250$\lambda^2$ for $\lambda \le 1250$ nm. Even with such AO systems, and apart from pure performance, multiplexing SM devices will be difficult in these conditions by necessitating a few hundreds such devices per spaxel, with its associated detector, or optimized versions to minimize the number of pixels, with cost on complexity and volume. For $\lambda \ge 1500$ nm, $S\omega$ is less than 15, and SM multiplexing can be considered as a viable solution. Note that multiplying the number of pixels by factor of 100 and more through SM multiplexing is not perceived as a fundamental issue in the context of ELTs: an instrument like HIRES/EELT will require slicing the input PSF with a fiber bundle of more than 200 fibers, choice imposed by the current production capabilities of bulk grating (private com.).

%%%%%%%%%%%%%%%%%%%%%%%%%%%%%
\begin{table}[t]
\centering
\caption{Parameters used for the performance estimations.}
\label{tab:sim_param}
\begin{tabular}{llrrr}
\hline \hline
Band && VIS (B) & NIR (H) & MIR (M) \\
\hline
Integration time  & [s]               & $10^3$ & $100$ & $0.01$ \\
RON  & [e-]           & 2 & 10 & 15 \\
Background & [ph/s] & $10$  & $10^3$ & $10^7$ \\
\hline
\end{tabular}
\end{table}
%%%%%%%%%%%%%%%%%%%%%%%%%%%%%

\paragraph{\bf Spatially resolved source}
\label{part:resolved_source}

Injecting light from a resolved source (whose size is bigger than the telescope limit of diffraction) into a SM waveguide is not optimal. It is difficult to estimate the coupling efficiency in this case as it depends on many parameters, among them, the intensity distribution of the source over the FOV of the fiber, the atmospheric turbulence strength $D/r_0$, the partial AO correction degree, etc. This study is out of the scope of this document but we can make an upper limit estimation of the coupling efficiency. Since SM waveguides can only accept a normalized \'etendue equal to 1, we consider at first order that a maximum of $1/S\omega$ of the flux can be coupled:
\begin{equation} \label{eq:rho_resolved}
	\rho < 1 / S \omega
\end{equation}
\noindent This relation can also be inferred in the case of an unresolved source in turbulent conditions. This simple assumption is sufficient for this study.

\subsubsection{{Coupling efficiency into M-SM and MM instruments}\label{part:coupling_efficiency_MM_MSM}}
Conversely to SM instruments, M-SM and MM ones benefit of an important optical \'etendue that can cover the source one, so that for them we consider $\rho \sim 1$. This is not true for few-modes instruments with M $<$ 50, with coupling that can decrease to $\rho \sim 60$\% for 30\% obscuration. I this study, this however represents a small adjustment, so we consider $\rho$ = 1 for MM and M-SM instruments.

%%%%%%%%%%%%%%%%%%%%%%%%%%%%%%%%%%%%%%%%%%%%%%%%%%%%%%%%
\begin{figure*}[b!]
  \centering
 \includegraphics[width=0.48\textwidth]{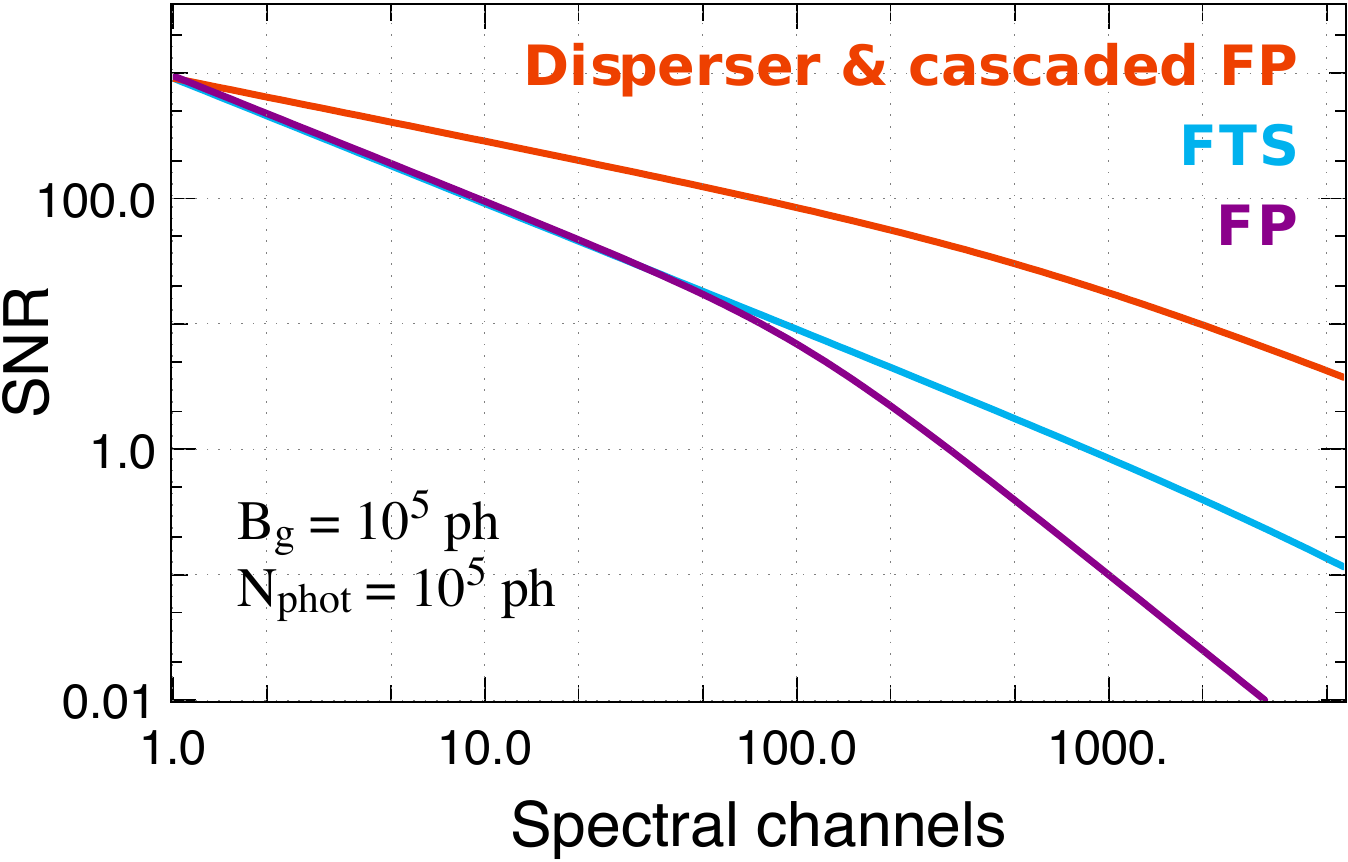}\hfill
 \includegraphics[width=0.49\textwidth]{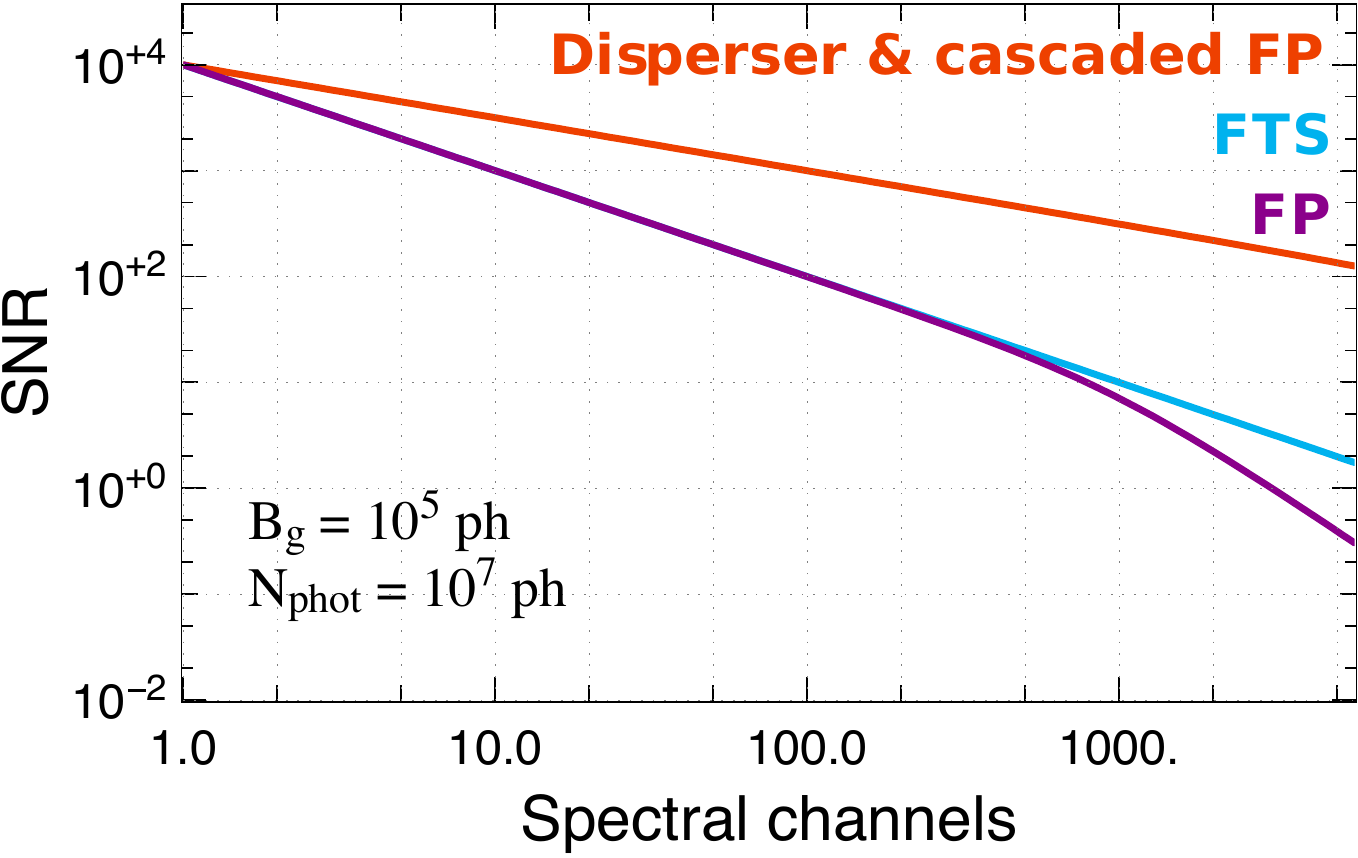}\\
   \caption[SNR vs spectral resolution in NIR.]{Expected SNR with respect to the number of spectral channels for the different families of spectrometers in NIR. $N_\lambda$ = 1000 corresponds to $\SR \sim 2000$ over the H band. {\it Left:} case of a source as luminous as the sky ($N_{phot}$ = $10^5$; $H_{mag}$ = 14.4); {\it Right:} case of a source 100 times more luminous than the sky background ($N_{phot}$ = $10^7$; $H_{mag}$ = 9.4).
  \label{fig:snr-vs-resolution-NIR}}
\end{figure*}
%%%%%%%%%%%%%%%%%%%%%%%%%%%%%%%%%%%%%%%%%%%%%%%%%%%%%%%%

\subsection{Performance as a function of the band} \label{part:perf_band}

We give a first insight on the relative performance of the different type of spectrographs in different bands, with a source of arbitrary extent and spectrographs adapted to this (i.e.~so that $\rho$ =1). Considerations on the source extent and observing condition is done in Section~\ref{part:extended_source}. The parameters for the comparison are given in Table~\ref{tab:sim_param}.

We consider two different sources. A faint object with {\it surface brightness} equal to the background, and a bright sources 100 times brighter. For an unresolved source, the energy is therefore spread over the instrument (turbulent) PSF. For bulk instruments, transmission of the different instruments can vary from 40\% for high resolution gratings to 90\% for FTS (including contrast losses). These numbers include only parts which are not shared by the spectrographs. For integrated spectrometers, the transmission depends on the technology rather than on the instrument family. The differences between the different possible solutions can slightly modify relative performances but, as we look at the global performance on a wide range of cases, they do not significantly change the conclusions. For the sake of clarity, we therefore consider that all spectrometers have a relative throughput of 1.

The signal-to-noise ratio of FTS, FP and Dispersers is compared in NIR in Fig.~\ref{fig:snr-vs-resolution-NIR} for two source brightness. The comparison over different bands shows that the SNR depends mainly on the number of spectral channels $\Nlambda$ and on the relative flux between the source and the background, rather than on absolute parameters like the bandwidth, spectral resolution, etc. The main difference between bands is the RON, which is negligible up to $\Nlambda\sim 10^3$ in IR and $\Nlambda\sim 10^4$ in visible. Dispersers and cascaded FP make the best use of photons, with SNR more than 20 times higher for$N_\lambda > 1000$.

Note that the reduced FTS performance at high resolution can be regained by reducing the transmitted bandwidth so that the number of spectral channels $\Nlambda$ is much smaller. A pre-dispersion with a prism or grating allows to keep the total bandwidth of interest by measuring several interferograms with a reduced bandwidth in parallel \cite{vaneyken_2010a}. The SNR of spectrographs computed here consider a source of arbitrary size in the sky, and spectrographs gathering the same quantity of light. In practice, it is easier to have access to a large FOV with scanning FTS or FP as demonstrated with instruments like TAURUS \cite{atherton_1982a} or still recently SITELLE \cite{drissen_2010a}.

%%%%%%%%%%%%%%%%%%%%%%%%%%%%%%%%%%%%%%%%%%%%%%%%%%%%%%%%%%%%%%%%%
\begin{figure*}[t]
\centering
\includegraphics[width=0.49\textwidth]{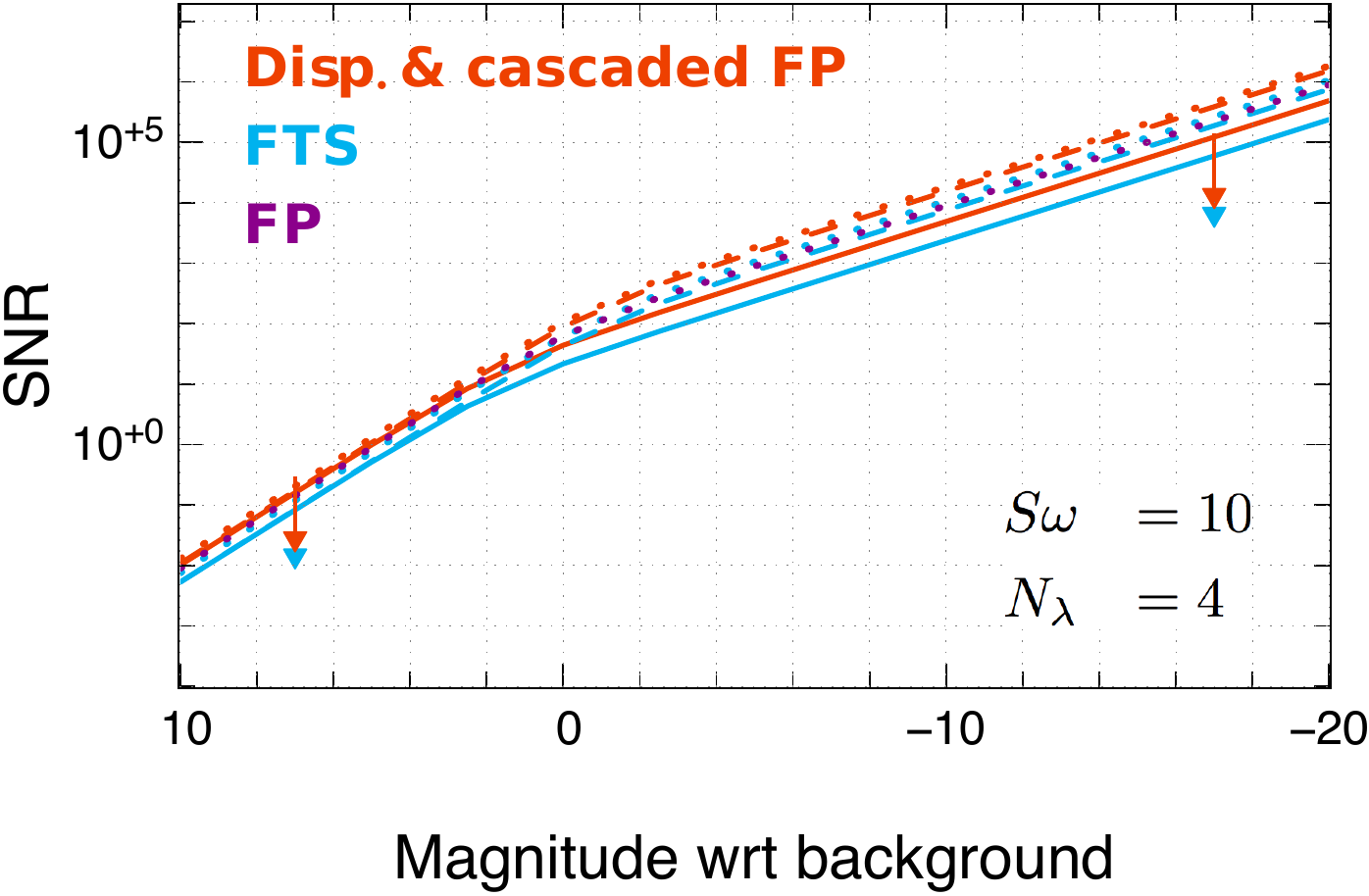}
\hfill
\includegraphics[width=0.49\textwidth]{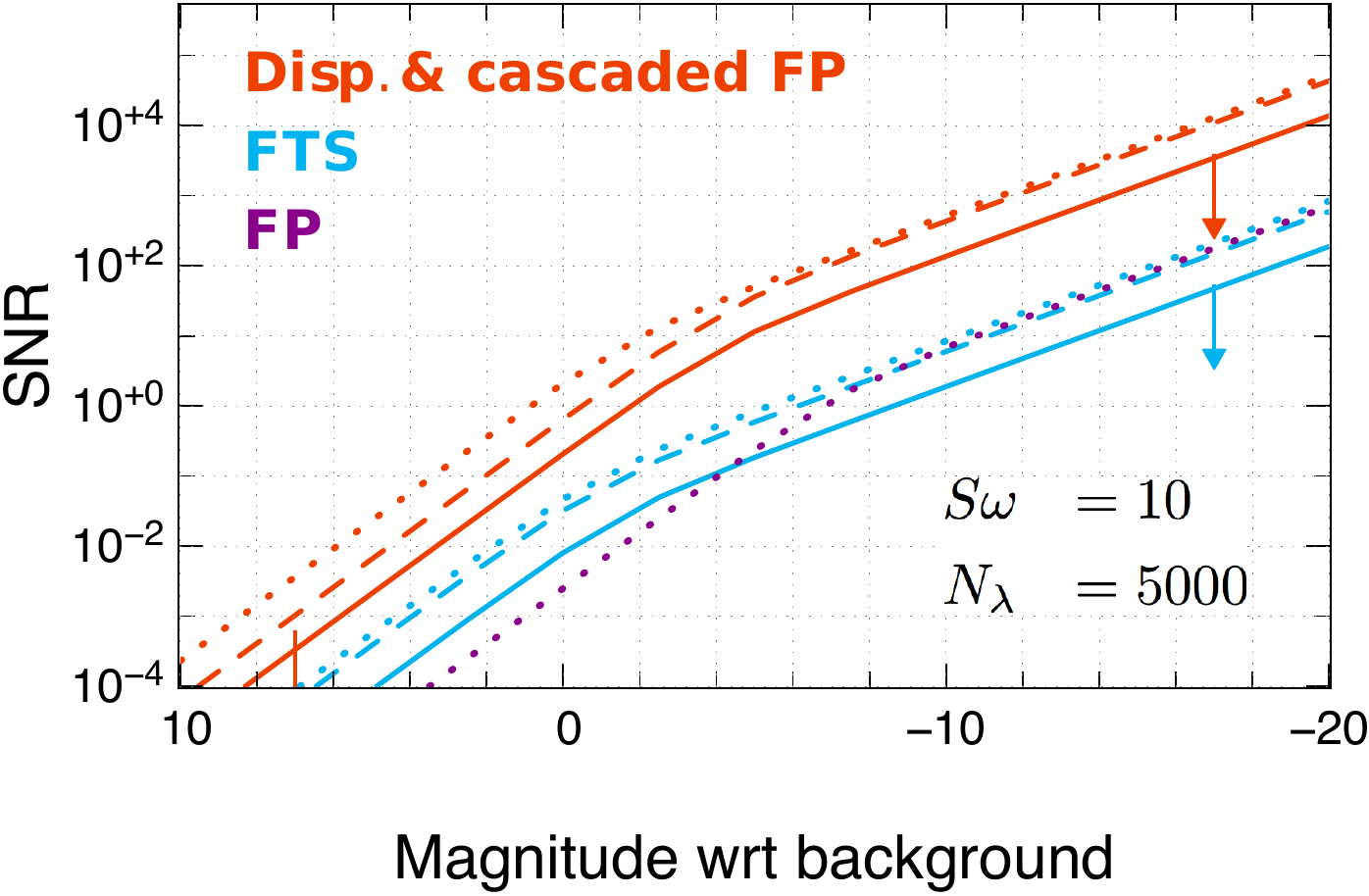}\\
\bigskip
\includegraphics[width=0.49\textwidth]{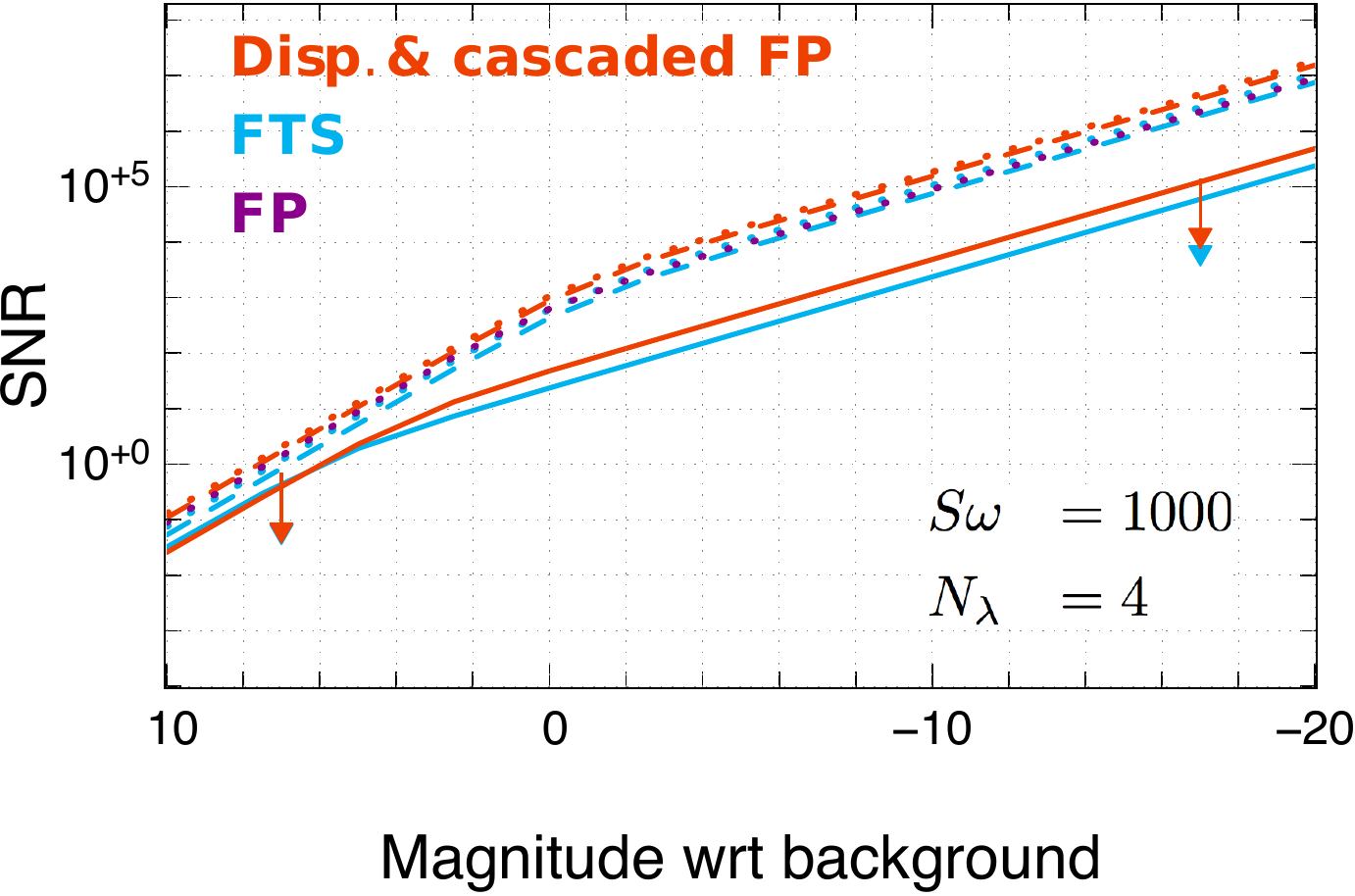}
\hfill
\includegraphics[width=0.49\textwidth]{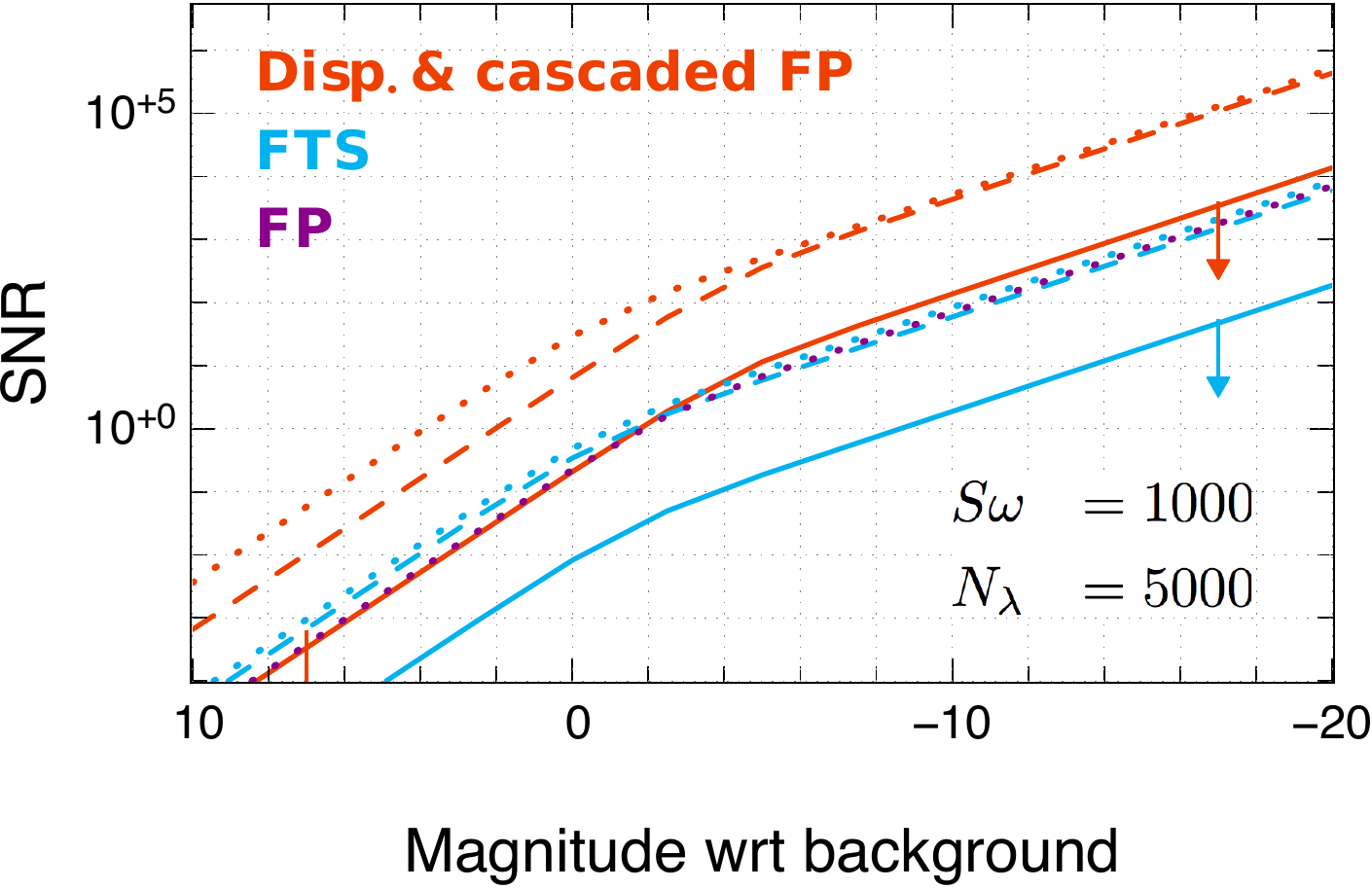}
\caption[SNR vs number of photons for a small source]{SNR vs number of photons for low and high number of spectral channels (left and right columns respectively). Top and bottom rows correspond to the small and extended source cases described in the text. We consider SM (solid), M-SM (dash) and MM (dotted) FTS, Dispersers and FP. MM curves are slightly shifted for the sake of clarity. \label{fig:SNRvsNphot}}
\end{figure*}
%%%%%%%%%%%%%%%%%%%%%%%%%%%%%%%%%%%%%%%%%%%%%%%%%%%%%%%%%%%%%%%%%

\subsection{Performance with an extended source} \label{part:extended_source}

We compare here the performance of the different types of spectrograph depending on number of spectral channels, the relative brightness of the sky and source, and on the source extent. We distinguish here 4 cases, crossing low and high number of spectral channels ($N_\lambda$ = 4 \& 5000) with sources of small and large optical \'etendue ($S\omega$ = 10 \& 1000). We recall that $S\omega$ = 1000 is a typical value for 50\% EE on an ELT in the NIR. From the result of previous section, we do not focus on a particular waveband, but rather look at the brightness of sky and source.

%%%%%%%%%%%%%%%%%%%%%%%%%%%%%%%%%%%%%%%%%%%%%%%%
\begin{figure}[b]
\centering
\includegraphics[width=0.95\textwidth]{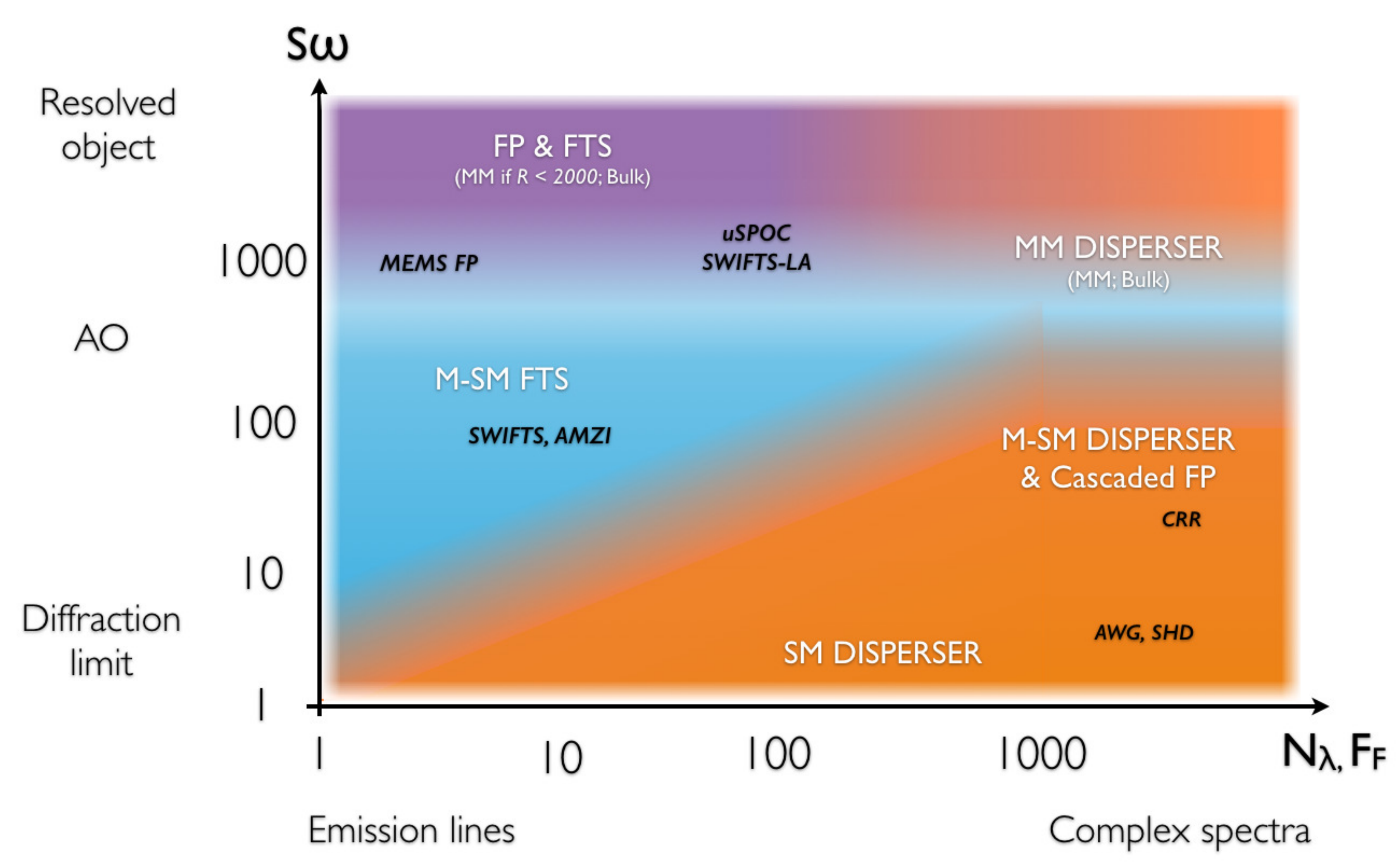}
\caption[Figure of merit of the different kind of instruments in the framework of astrophotonics.]{Figure of merit of the different kind of instruments in the framework of astrophotonics. Following the conclusions of the performance analysis, this figure shows the preferred operating regimes of the different families of spectrographs, depending on the spectrum complexity (linked to the number of spectral channels) and on the source extent. The different areas follow the nomenclature for the performance plots: orange for dispersers, blue (and violet) for FTS, and violet for FP.  We also report in black the identified technologies with the best performances. \label{fig:SpectroSumUp}}
\end{figure}
%%%%%%%%%%%%%%%%%%%%%%%%%%%%%%%%%%%%%%%%%%%%%%%%

\begin{itemize}
\item For small sources, dispersers and cascaded FP are the best compromise, whatever their type [Fig.~\ref{fig:SNRvsNphot} (top)]. For bright sources and low spectral resolution, M-SM FTS and bulk FP are slightly better than SM dispersers. The advantage goes up to $\Nlambda \sim 10$, but the gain is not more than a factor of 2 in SNR. At high spectral resolution dispersers and cascaded FP can be more sensitive by at least 3 magnitudes.
\item For $S\omega > 1000$ (most cases in visible for an ELT), M-SM FTS and FP are slightly as efficient as a single SM disperser, whatever the flux and the number of spectral channels. At low number, SM dispersers become less efficient by a factor 100. For a number of spectral channels bulk FP performance quickly degrade for $\Nlambda > 5000$.
\end{itemize}

If we put those results in the framework of the ELTs science cases, which aim at important number of spectral channel per spaxel, and considering the AO performance, most cases are situated on figure on the bottom right panel of Fig.~\ref{fig:SNRvsNphot}. We should bare in mind that the coupling of the SM disperser in such conditions is $\sim 10^{-3}$ of what collects an ELT. In other words, a few 10cm telescope would perform as well! So, in most cases, FTS (bulk or integrated) and bulk FP would make a pretty bad use of the surface offered by an ELT, and effort should tend towards developing  dispersive and cascaded FP solutions, in combination with energy sensitive detectors.

 % % % % % % % % % % % % % % % % % % % % % % % % % % % % % % % % % % % % % % % % % % % %
 \section{Conclusion} \label{part:conclusion}

The goal of this study was to compare the performance of the different possible integrated spectrographs, in a framework as general as possible, and to present a non-exhaustive list of integrated spectrograph technologies of potential interest for astrophotonics. Fig.~\ref{fig:SpectroSumUp} gives a summary of this work by showing the preferred operating regimes of the different concepts presented here, and in which conditions the best identified technologies could be used. 

Efforts are already under way to develop and demonstrate the potential of integrated spectrographs in astrophysics, e.g.~with AWGs which have begun to be tested on sky \cite{cvetojevic_2012c}. We emphasize in this work that Cascaded Ring Resonator is an equally promising solution, although no work for astrophotonic spectrographs is underway. In this context, developments of AO systems in the coming years will be fundamental in limiting the level of multiplexing at all levels, especially regarding spectrographs and detectors. A solution of growing interest resides in the development of few-mode spectrographs: they could deliver spectral resolutions of a few 1000 while increasing \'etendue to $S\omega \sim 50$ \cite{horton_2007a, corbett_2009a}. This should also go with the development of energy-sensitive detectors (and polarization-sensitive for $\SR \ge$ 10000), in order to build fully integrated spectrographs, while preserving compactness and efficiency. 

As a final word, this study paves the way to future works addressing technological, implementation and exploitation issues of the presented spectrometers in a more specific framework. Such a study is currently carried out in the framework of planetology and dedicated space missions, especially considering imaging spectrometry [S. Gousset, E. Le Coarer, B. Schmitt, in prep.]. The comparison metric presented in this paper will be first applied then completed by taking into account both instrumental complexity and data processing relevant to each reviewed technology.

\section*{Funding}
LabEx FOCUS ANR-11-LABX-0013; OPTICON Astrophotonics (Work Package 3)

\section*{Acknowledgments}
We thank the two referees for their comments that helped clarifying the content of this paper.

\end{document}